\documentclass[aps,showpacs,jcp,superscriptaddress,floatfix,twocolumn,citeautoscript]{revtex4-2}
\usepackage{graphicx}
\usepackage{amsmath}
\usepackage{amssymb}
\usepackage{float}
\usepackage{bm}
\usepackage{dcolumn}
\usepackage{hyperref}
\usepackage[usenames]{color}
\usepackage{multirow}
\hypersetup{pdfborder=0 0 0,colorlinks=true,citecolor=blue,linkcolor=blue}
\input epsf
\DeclareGraphicsExtensions{.pdf,.jpeg,.png, .eps}
\usepackage[labelsep=period,justification=raggedright,singlelinecheck=false,figurename=Figure]{caption}

\begin{document}
	\title{Coulomb drag in metal monochalcogenides double-layer structures with Mexican-hat band dispersions}
	
		\author{S. Rostami}
		\affiliation{Department of Physics, Shahid Beheshti University, G. C., Evin, Tehran 1983969411, Iran}
		\author{T. Vazifehshenas}
		\email{t-vazifeh@sbu.ac.ir}
		\affiliation{Department of Physics, Shahid Beheshti University, G. C., Evin, Tehran 1983969411, Iran}
        \author{T. Salavati-fard}
	    \affiliation{Department of Chemical and Biomolecular Engineering, University of Houston, Houston, Texas 77204, United States}
    	\affiliation{Texas Center for Superconductivity at the University of Houston, Houston, Texas 77204, United States}
	\begin{abstract}
		
	We theoretically study the Coulomb drag resistivity and plasmon modes behavior for a system composed of two parallel p-type doped GaS monolayers with Mexican-hat valence energy band using the Boltzmann transport theory formalism.
	We investigate the effect of temperature,$\ T$, carrier density,$\ p$, and layer separation,$\ d$, on the plasmon modes and drag resistivity within the energy-independent scattering time approximation. Our results show that the density dependence of plasmon modes can be approximated by$\ p^{0.5}$. Also, the calculations suggest a$\ d^{0.2}$ and a$\ d^{0.1}$ dependencies for the acoustic and optical plasmon energies, respectively. 
	Interestingly, we obtain that the behavior of drag resistivity in the double-layer metal monochalcogenides swings between the behavior of a double-quantum well system with parabolic dispersion and that of a double-quantum wire structure with a large carrier density of states. In particular, the transresistivity value reduces exponentially with increasing the distance between layers. Furthermore, the drag resistivity changes as$\ T^{2}/p^{4}$ ($\ T^{2.8}/p^{4.5}$) at low (intermediate) temperatures. Finally, we compare the drag resistivity as a function of temperature for GaS with other Mexican-hat materials including GaSe and InSe and find that it adopts higher values when the metal monochalcogenide has smaller Mexican-hat height.			
\end{abstract}
	
	
	
	\maketitle
	
	\section{Introduction}
	Two-dimensional (2D) materials have been among most extensively studied structures due to the wide range of applications in nanoscience and nanotechnology \cite{xie2015two,fiori2014electronics,wu2018thermo,zhang2015strain}. Recently, the atomically thin layers of metal monochalcogenides as a new class of this family, has attracted much attention \cite{demirci2017structural,zhou2017multiband,cao2015tunable,shi2015anisotropic}. These 2D materials have special electronic and structural properties which make them promising candidates for different applications such as field-effect transistors(FETs), electronic sensors, and solar energy and photoelectric devices \cite{cai2019synthesis,feng2015performance,wang2015high,wang2019first,budweg2019control}.
	The general chemical formula of these layered materials is
	MX, where M belongs to Group III and X refers to Group VI in the table of elements. GaS, GaSe and InSe are some examples. In their bulk form,  there is a strong covalent chemical bond between metal and  chalcogenide atoms and each layer is coupled to its neighboring layers by the van der Waals forces \cite{hu2015gese,bejani2019lattice,ariapour2020strain}. When the thickness of this group of materials reduces to few monolayers, their valance band looks like a “Mexican-hat” \cite{seixas2016multiferroic}. A Mexican-hat dispersion forms ring-shaped valence band edges, at which the van Hove singularities appear with $\ 1/\sqrt {E}$ divergence in the 2D density of states (DOS) \cite{wickramaratne2015electronic,stauber2007fermi}. Exploring this novel class of 2D semiconductors with a large DOS near the Fermi surface, tunable magnetism, superior flexibility and good ambient stability is an important research topic in recent years.  
	In addition, successfully synthesizing monolayer and few-layer MXs, including GaS, GaSe and InSe, presents an intriguing opportunities for future semiconductor technology \cite{hu2012synthesis,zhou2018inse,chang2018synthesis,lei2013synthesis}.
	
	Over the course of past few decades, a great deal of attention has focused on double-layer 2D structures because of their interesting many-body and transport features which arise from the inter-layer Coulomb interaction between the two parallel electron or hole systems that are coupled in close proximity \cite{vazifehshenas2012thickness,tanatar2001dynamic,gumbs2018effect,perali2013high,vazifehshenas2015geometrical}. The Coulomb drag phenomenon provides an opportunity to  measure the effects of electron-electron interactions through the transport measurement, directly where the momentum is transferred from one  layer to the other layer due to the inter-layer Coulomb coupling \cite{hwang2011coulomb,narozhny2016coulomb,carrega2012theory,vazifehshenas2007thickness}. A driving current$\ (I_{drive})$ in one layer ("the active layer") induces a voltage$\ (V_{drag})$ in the other layer ("the passive layer"). This phenomenon is called Coulomb drag. The transresistivity or the drag coefficient$\ (\rho_{D})$  is a measure
	of inter-layer interaction and can be determined by calculating the ratio of$\ V_{drag}$ to$\ I_{drive}$ \cite{narozhny2012coulomb,sivan1992coupled}. This phenomenon has previously been studied in some nanostructures such as n-doped and p-doped double quantum wells \cite{flensberg1994coulomb,yurtsever2003many,hwang2003frictional,pillarisetty2005coulomb}, double quantum wires \cite{tanatar1998disorder,tanatar1996coulomb,tanatar2000effects}, mismatched subsystems \cite{badalyan2020coulomb}, double layers of topological materials \cite{liu2019coulomb}, double-layer and bilayer graphene \cite{tse2007theory,narozhny2012coulomb,hwang2011coulomb},
	and double-layer phosphorene \cite{saberi2016coulomb}.  For a double quantum wells system with a 2D electron density$\ n$ and layer separation$\ d$, the drag transresisitivity changes as$\ T^{2}/n^{2}d^{4}$ ($\ 1/Tn^{3(4)}d^{3}$) at low (high) temperature$\ (T)$. In the case of double-layer graphene with linear energy band dispersion, it has been found that$\ \rho_{D}$ has a $\ T^{2}/n^{2}d^{2}$  ($\ T^{2}/n^{4}d^{6}$) dependency at low (high) carrier density, while $\ \rho_{D}$ for a system of double graphene bilayers with quadratic dispersion shows a$\ T^{2}/n^{3}d^{4}$  ($\ T^{2}/n^{3}ln(d)$) behavior in the large (small) layer separation case \cite{hwang2011coulomb}. Also, at low (high) temperature, a system of double quantum wires exhibits a$\ T^{2}$($\ T^{-3/2}$) dependence within the Fermi liquid approach \cite{glazman2006coulomb}. 
	However, the Coulomb drag effect has not been studied for materials with the Mexican-hat dispersion and these interesting double-layer systems are still open for investigations. In MX monolayers, the Mexican-hat dispersion results in a high density of states and a van Hove singularity near the valence band maximum which can affect their electronic \cite{demirci2017structural,zhao2019magnetism}, optoelectronic \cite{magorrian2017spin,lei2016surface}, thermoelectric\cite{nurhuda2020thermoelectric,wickramaratne2015electronic,wang2019strain} and many-body properties. This motivates us to theoretically investigate the many-body Coulomb drag effect of such 2D materials with Mexican-hat band structure. Among above mentioned monolayer MXs, GaS has a larger Mexican-hat that can be attributed to the charge transfer, caused by the elements' electronegativities difference (Se$\ <$ S, In$\ <$ Ga), which occupies the $p$ orbitals of S or Se and dominates the top valence bands \cite{wang2019first}.
	
	In this paper, we theoretically investigate the Coulomb drag effect between two p-type doped identical parallel monolayers of a few III-VI compounds whose valance bands look like Mexican-hat. Special attention is paid to GaS with a lattice constant of$\ a = 3.46$ {\AA} which is known to have promising electronic and optical characteristics \cite{demirci2017structural,yagmurcukardes2016mechanical,ho2006optical,budweg2019control}. 
  We will start off with the expression for drag resistivity based upon the semiclassical Boltzmann transport equation and energy-independent scattering time approximation. Then, we will use a general formalism for calculating the drag resistivity in our desired system and the effects of various parameters such as temperature($\ T$), hole density ($p$) and layer separation($\ d$) will be investigated. In order to better understand the drag resistivity behavior, we also extract the double-layer plasmon modes as functions of the studied parameters from the dynamical dielectric function. We will finally present a comparison between drag resistivity of GaS monolayer and its some other family members such as GaSe and InSe monolayers.
  We have ignored the virtual phonon exchange effects on the drag transresistivity\cite{gramila1993evidence,tso1992direct} in our calculations. This mechanism is expected to be relevant at very low temperatures (where the contribution of plasmons to the drag is negligible) and for the large inter-layer separations (where the Coulomb interaction between the layers is weak)\cite{amorim2012coulomb,zarenia2019coulomb,flensberg1994coulomb}. In this study, the distance between two layers is chosen to be small (15-30{\AA}). Therefore, the Coulomb interaction between the layers is strong enough that one can safely neglect the effect of virtual phonon exchange. Also, the coupling between the plasmons and surface optical phonons of substrate is not taken into account because for most parameters used here, the Fermi energy, as a result of the large density of states at band edge (van Hove singularity), is very small and far below the surface optical phonon energy so that the interaction between them is almost negligible.

	The rest of this paper is structured as follows: in Sec. II, we describe the model and theoretical formalism. In Sec. III, we present results together with detailed discussion and finally, conclusion is given in Sec. IV.
	\section{Model and Formalism}
\begin{figure}[t!]
		\includegraphics[width=4cm]{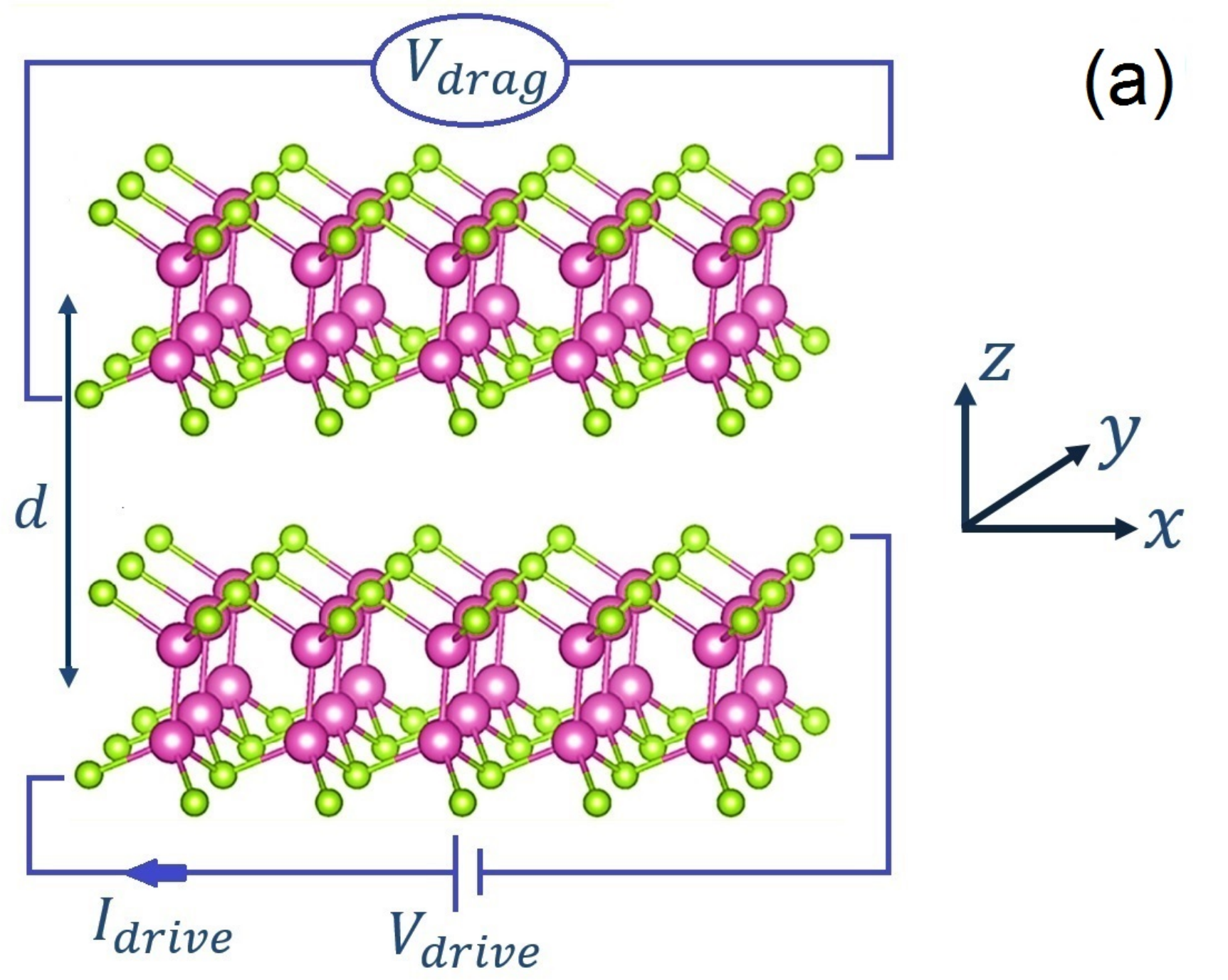}\includegraphics[width=4cm]{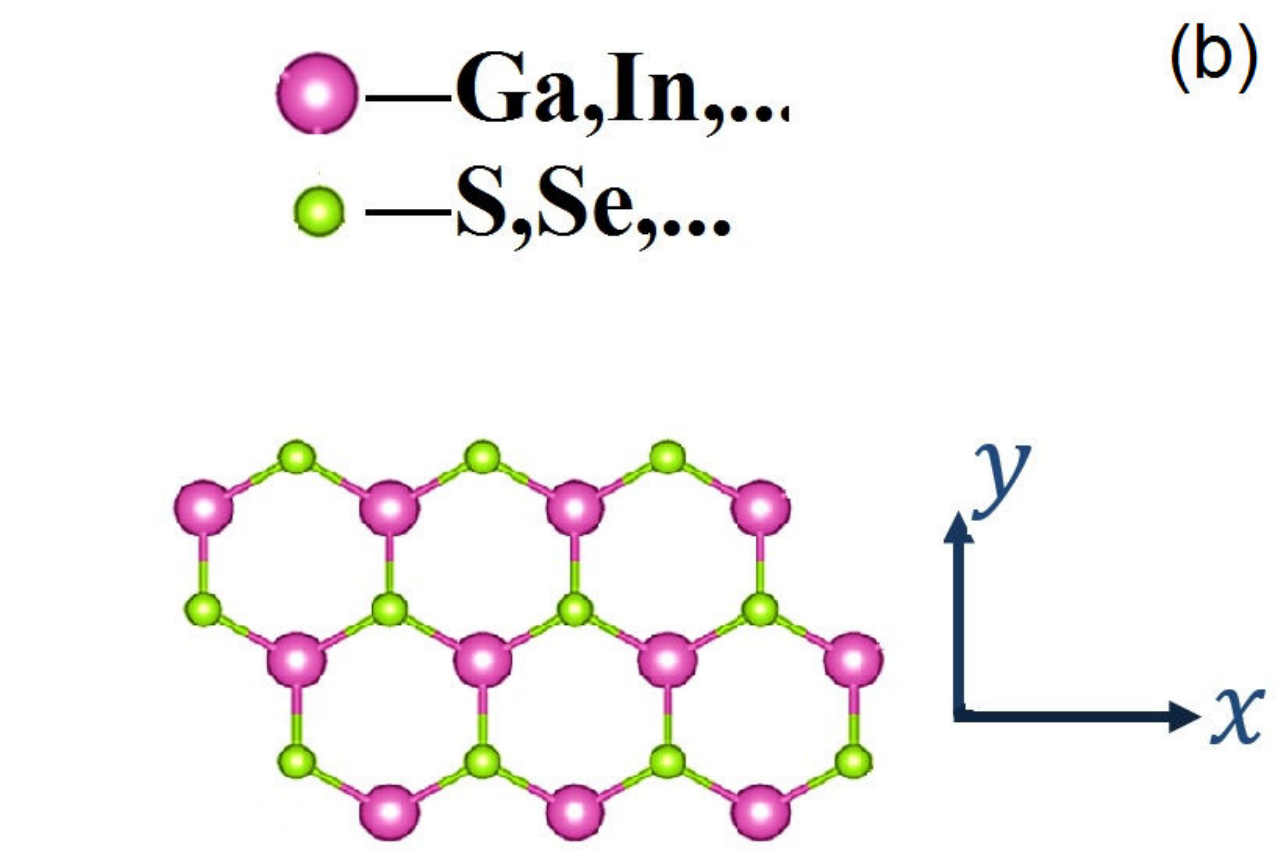}
		\caption{(a) Side view of a double-layer structure composed of III-VI compounds monolayers in a drag setup. (b) Top view of III-VI compounds general atomic structure.}\label{Fig-1:sil}
	\end{figure}

	The structure is modeled as two p-type doped identical parallel monolayers with Mexican-hat valence band dispersion which are coupled by Coulomb interaction in a short distance. The separation is still far enough to prohibit any electron tunneling. Figure \ref{Fig-1:sil}(a) shows a schematic model of this system and Figure \ref{Fig-1:sil}(b) demonstrates the top view of the crystal structure of III-VI compounds.
	The valence band energy dispersion relation of each layer is given by \cite{das2019charged}:
	\begin{equation}\label{eq:E}
	E(k)=E_{0} -\lambda_{1}k^{2} +\lambda_{2}k^{4}
	\end{equation}
	where $E_{0} $ is the height of the hat at$\ k=0 $ (see Figure \ref{Fig-2:mx}), $\lambda_{1}=\hbar^2/2m^{*}$, $\lambda_{2}=\hbar^4/4E_{0} {m^{*}}^{2}$
	and$\ m^*$ is  the hole effective mass at$\ k=0 $. $\ E_{0}$ and$\ m^{*}$ are, respectively, set to 111.2 meV and 0.409$\ m_{0}$ for GaS monolayer, with $\ m_{0}$ being the free electron mass \cite{wickramaratne2015electronic}.
	As shown in Figure \ref{Fig-2:mx}, the hole kinetic energy is assumed to be positive. According to the dispersion energy equation given above, the valence band edge is located at$\ E = 0$ and negative energies represent energies in the bandgap. There are two Fermi wave vectors,$\ {{k}_{F}}_{1}$ and$\ {{k}_{F}}_{2}$, for positive Fermi energies smaller than$\ E_{0}$ in the Mexican-hat dispersion. These two Fermi wave vectors originate from the two branches of the dispersion with concentric ring radii of$\ {k_{F}}_{1}=\sqrt{(4m^{*}E_{0}/\hbar^{2})(1-\sqrt{E/E_{0}})}$ and$\ {k_{F}}_{2}=\sqrt{(4m^{*}E_{0}/\hbar^{2})(1+\sqrt{E/E_{0}})}$ corresponding to the Fermi surface. Density of states for a 2D Mexican-hat structure is given by \cite{das2019charged}:
	\begin{figure}[ht!]
	\includegraphics[width=6cm]{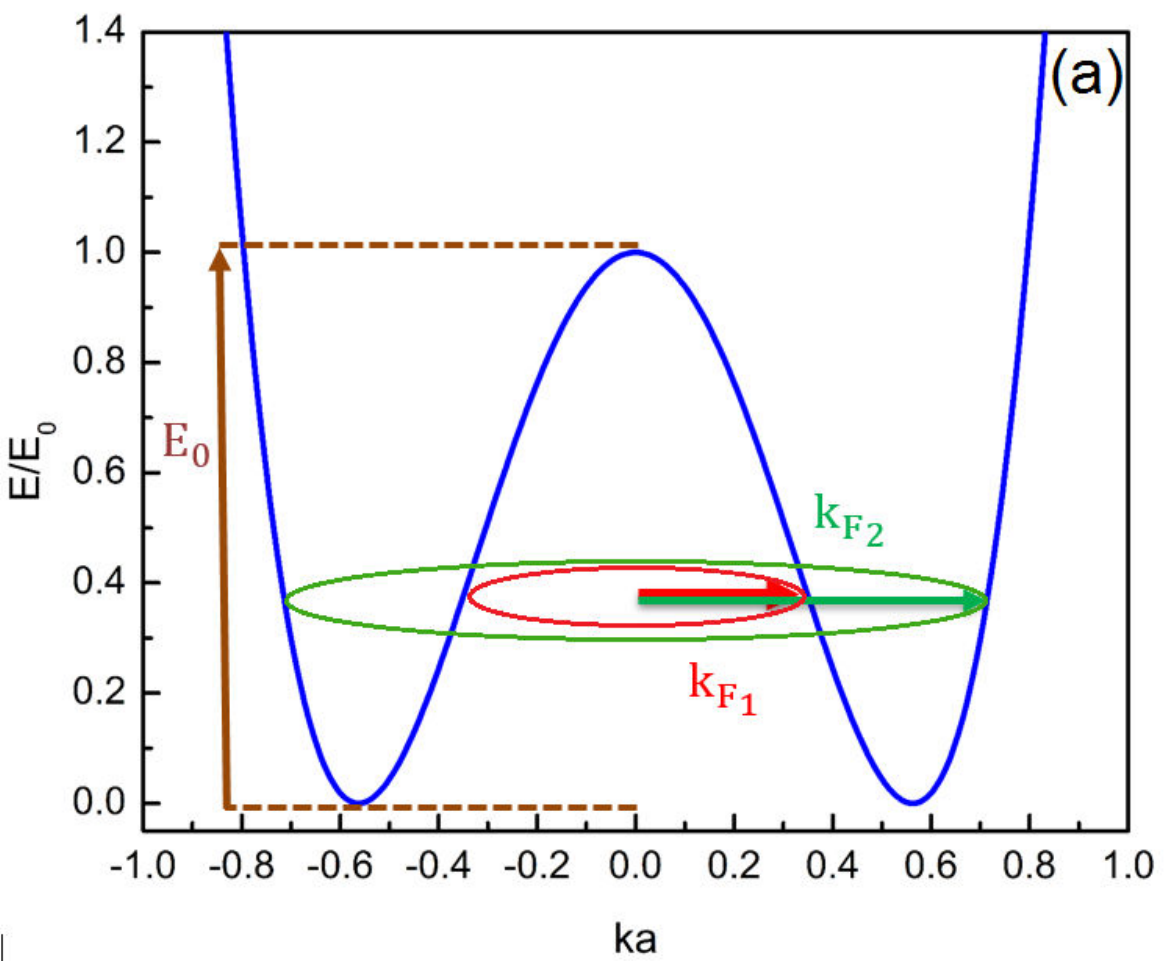}
	\includegraphics[width=4cm]{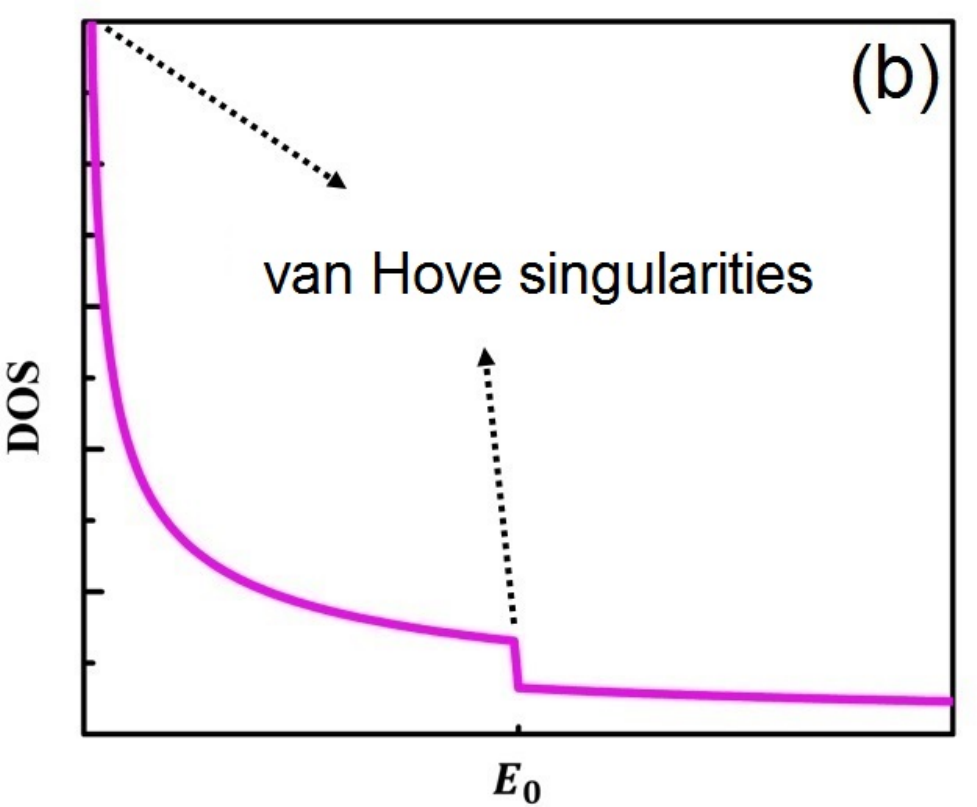}
	\caption{(a) Mexican-hat dispersion for GaS monolayer with$\ p=5\times10^{13} $cm$^{-2}$ and$\ E_{F}=0.37E_{0}$. The two concentric rings show the two Fermi circles with radii $\ {k_{F}}_{1}$ and$\ {k_{F}}_{2}$ that exist at Fermi energies below$\ E_{0}$. (b) DOS and van Hove singularities.}\label{Fig-2:mx}
	\end{figure}
	\begin{equation}\ \label{eq:DS}
	DOS(E) = \left\{
	\begin{array}{rl}
	\frac{2m^*}{\pi \hbar^2} \sqrt{\frac{E_{0}}{E}}  & \ \ \ E<E_{0} \\
	\frac{m^*}{\pi \hbar^2} \sqrt{\frac{E_{0}}{E}}   & \ \ \ E>E_{0}
	\end{array}\right.
	\end{equation}
	with the Fermi energy $\ E_F=p^{2}\pi^{2}\hbar^{4}/16E_{0} {m^{*}}^{2} $
	where $p$ is the 2D hole density. The Mexican-hat electronic band structure leads to divergences in the density of states, the so-called van Hove singularities: the first one diverges with$\ 1/\sqrt{E} $ behavior at$\ E = 0$ and another is a Heaviside step function discontinuity at$\ E = E_{0}$. Existence of the van Hove singularities promises new electronic properties when the Fermi energy is in close vicinity \cite{zhou2017multiband,rybkovskiy2014transition}.
	
	The drag conductivity is defined by:
	\begin{equation}\label{eq:sigma}
	\sigma_D=\frac{J^\alpha_1}{E^\alpha_2}
	\end{equation}
	where$\ \alpha$ is the direction along $\ x$ or $\ y$ in which the current $\ J_1$ flows. Indices 1 and 2 denote the active and passive layers, respectively. The drag resistivity relates to the layers conductivities in isotropic systems as follows :
	\begin{equation}\label{eq:rh} 
	\rho_{D}\simeq-\frac{\sigma_D}{\sigma_{11}\sigma_{22}}
	\end{equation}
	where $\sigma_{11}$  and $\sigma_{22}$ are the intra-layer conductivities of the active and passive layers, respectively. The drag resistivity can be obtained through several methods such as the
	Kubo formula based on the leading-order diagrammatic perturbation theory\cite{flensberg1995linear,kamenev1995coulomb},  the memory function formalism\cite{zheng1993coulomb} and
	the linear response Boltzmann transport equation\cite{jauho1993coulomb}. Within the third approach, the drag resistivity is given by
	\cite{flensberg1994coulomb}:
	\begin{equation}\label{eq:rhoD}
	\begin{split}
	\rho_{D}=-\frac{{m^{*}}_{1}{m^{*}}_{2}}{4\pi k_{B}T p_{1}p_{2} e^{4}\tau_{1}\tau_{2}} \qquad\qquad\\ \times\sum_{\textbf{q}}\int d\omega \frac{Im[\Gamma^{\alpha}_{1}(\textbf{q},\omega)]Im[\Gamma^{\alpha}_{2}(\textbf{q},\omega)] {\left|W_{12}(q,\omega)\right|}^2}{sinh^2(\hbar\omega/2k_{B}T)}.
	\end{split}	
	\end{equation}
	\ $\omega$ and $\ \textbf{q}$ are the transferred energy and momentum from layer 1 to layer 2 at temperature T, $\ k_{B}$ is the Boltzmann constant,$\ W(\textbf{q},\omega)$ is the screened inter-layer Coulomb interaction and$\ \tau_{1(2)}$ is the transport scattering time of layer 1 (layer 2). We assume that the relaxation time is not energy dependent and both layer 1 and layer 2 are identical with equal hole densities. $\ \Gamma^{\alpha}_{i}(\textbf{q},\omega)$ is the non-linear susceptibility along $\alpha$ direction which is given as \cite{zheng1993coulomb}:
	\begin{equation}\label{eq:Fi}
	\vspace{2mm}
	\Gamma^{\alpha}_{i}(\textbf{q},\omega)=g\sum_{\textbf{k}}\frac{e(f_{i}(\textbf{k})-f_{i}(\textbf{k}^{'}))(\tau_{i}v^{\alpha}(\textbf{k})-\tau_{i}v^{\alpha}(\textbf{k}^{'}))}{E(\textbf{k})-E(\textbf{k}^{'})+\omega+i\eta^{+}}.
	\end{equation}
	In this equation$\ \textbf{k}{'}=\textbf{k}+\textbf{q}$,$\ g$ is spin degeneracy,$\ v^{\alpha}(\textbf{k}) $ is the $\alpha$ component of group velocity ,$\ e$ is the electron charge and$\ f(\textbf{k})=\{exp[(E(\textbf{k})-\mu)/k_{B}T]+1\}^{-1}$ is the equilibrium Fermi distribution function with $\mu$ being the chemical potential. $\ E(\textbf{k})$ refers to the Mexican-hat dispersion given in Eq. (\ref{eq:E}). The 2D non-linear susceptibility in a special direction such as $\ x$ can be obtained as:
	\begin{equation}\label{eq:Gama}
	\Gamma^{x}_{i}(\textbf{q},\omega)=\sum_{\textbf{k}}\dfrac{e\tau [f_{i}(\textbf{k})-f_{i}(\textbf{k}^{'})]{\Delta v^{x}_{\textbf{k},\textbf{k}^{'}}}}{\Delta E_{\textbf{k},\textbf{k}^{'}}+\omega+i\eta^{+}}
	\end{equation}
	where $\ {\Delta v^{x}_{\textbf{k},\textbf{k}^{'}}}$ and $\ \Delta E_{\textbf{k},\textbf{k}^{'}}$ are given by following relations:
	\begin{equation}\label{eq:deltaV}
	\begin{split}
	{\Delta v^{x} _{\textbf{k},\textbf{k}^{'}}}=\frac{2\lambda_{1}q_{x}}{\hbar}+\frac{4\lambda_{2}}{\hbar} [{k_{x}}^3-{(k_x+q_x)}^3\\+k_{x}{k_{y}}^2-(k_{x}+q_{x}){(k_{y}+q_{y})}^2].
	\end{split}
	\end{equation}
	\vspace{0.2cm}
	and
	\begin{equation}\label{eq:deltaE}
	\Delta E_{\textbf{k},\textbf{k}^{'}}=A(k,q)cos^{2}\theta+B(k,q)cos\theta+C(k,q)
	\end{equation}
	with A, B and C defined as
	\begin{equation}\label{eq:A}
	A(k,q)=4\lambda_{2} k^2 q^2
	\end{equation}
	\begin{equation}\label{eq:B}
	B(k,q)=2 \lambda_{1}kq-4\lambda_{2}k q^3-4\lambda_{2}k^3 q
	\end{equation}
	\begin{equation}\label{eq:C}
	C(k,q)=2 \lambda_{2} k^2 q^2 +\lambda_{2}q^4-\lambda_{1}q^2
	\end{equation}
	where$\ \theta$ is the angle between$\ \textbf{k}$ and$\ \textbf{q}$.\\
	
	\section{Results and discussion}
	    \begin{figure*}[h]
		\centering
		\includegraphics[width=7cm]{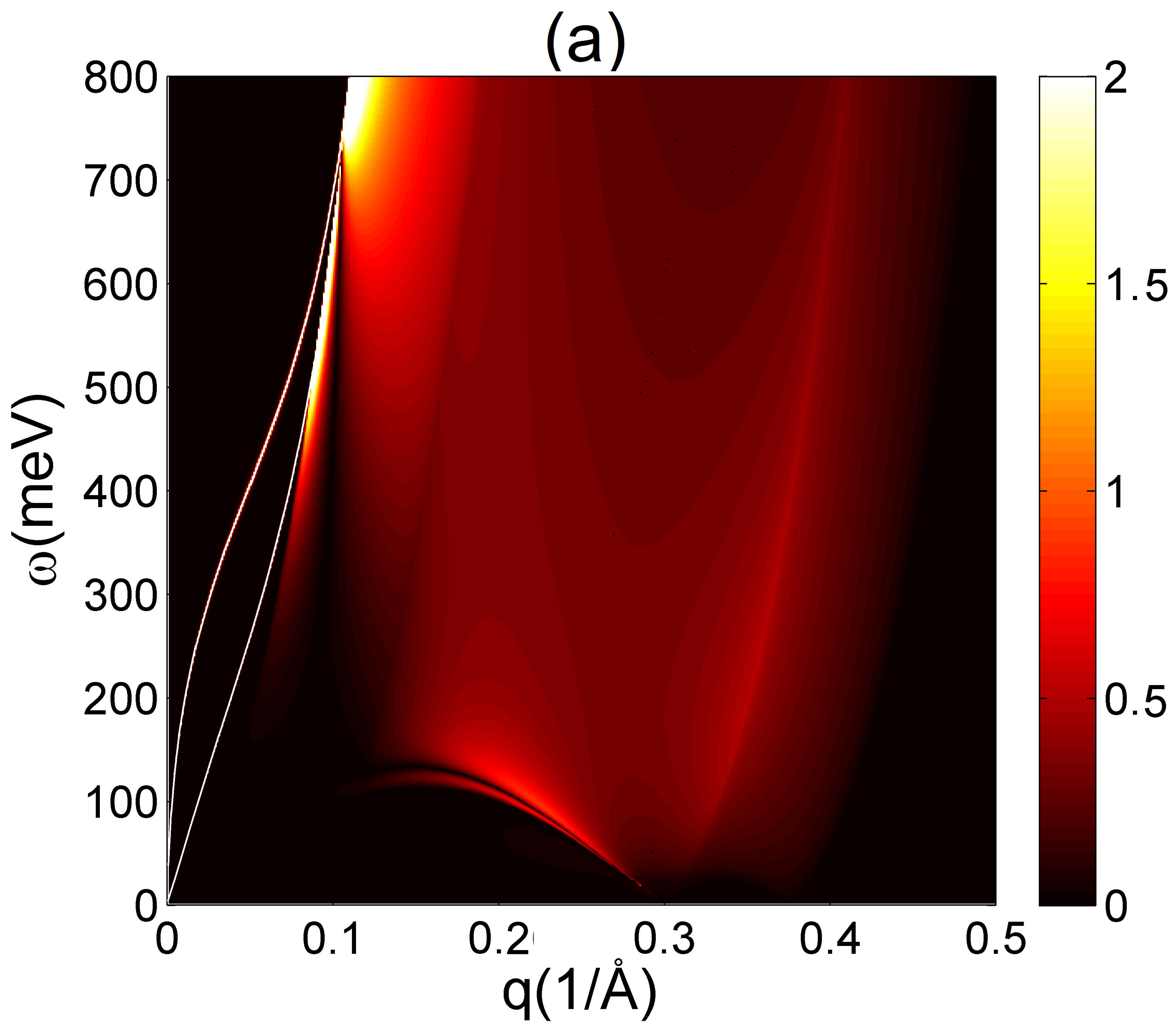}
		\hspace{0.5cm}
		\includegraphics[width=7cm]{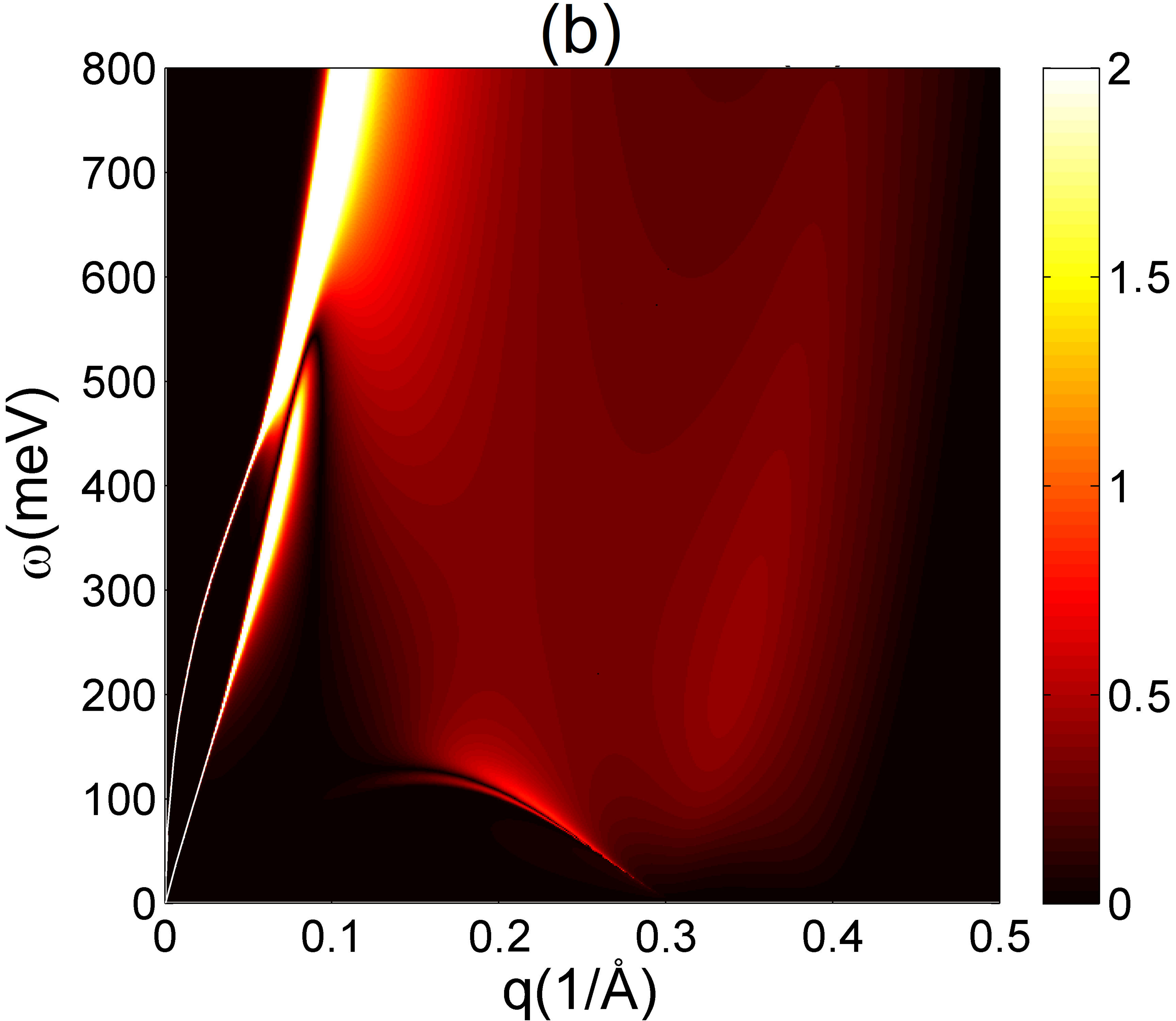}
		\caption{The loss function for a double-layer structure of GaS monolayers for$\ d=15$ {\AA} and$\ p=4\times10^{13} $cm$^{-2} $ at two temperatures: (a)$\ T=0$  and (b)$\ T=0.5 T_{F}$ .}\label{Fig-3:loss3}
    	\end{figure*}
    \begin{figure*}[h]
	\centering
	\includegraphics[width=5.6cm]{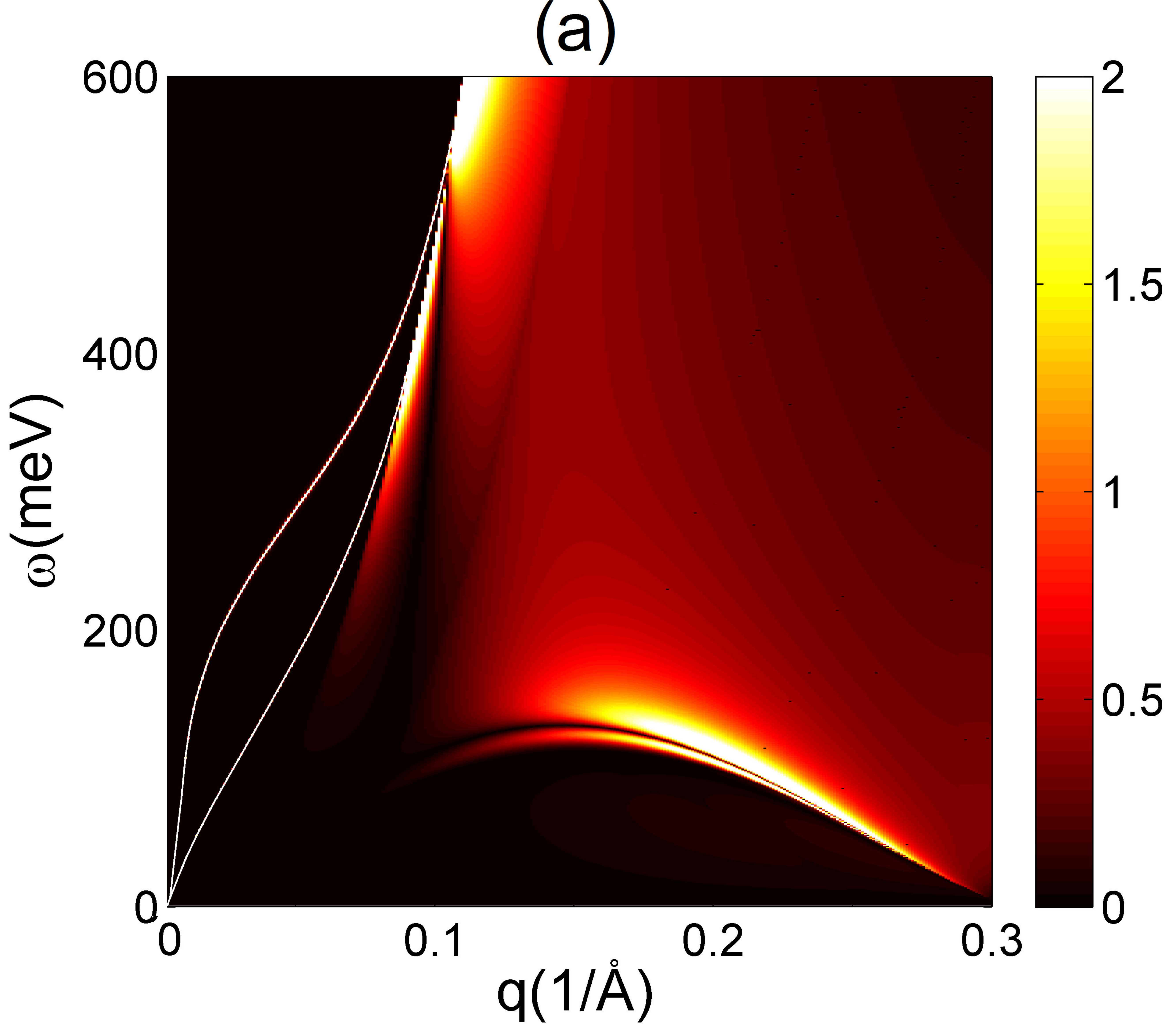}
	\includegraphics[width=5.6cm]{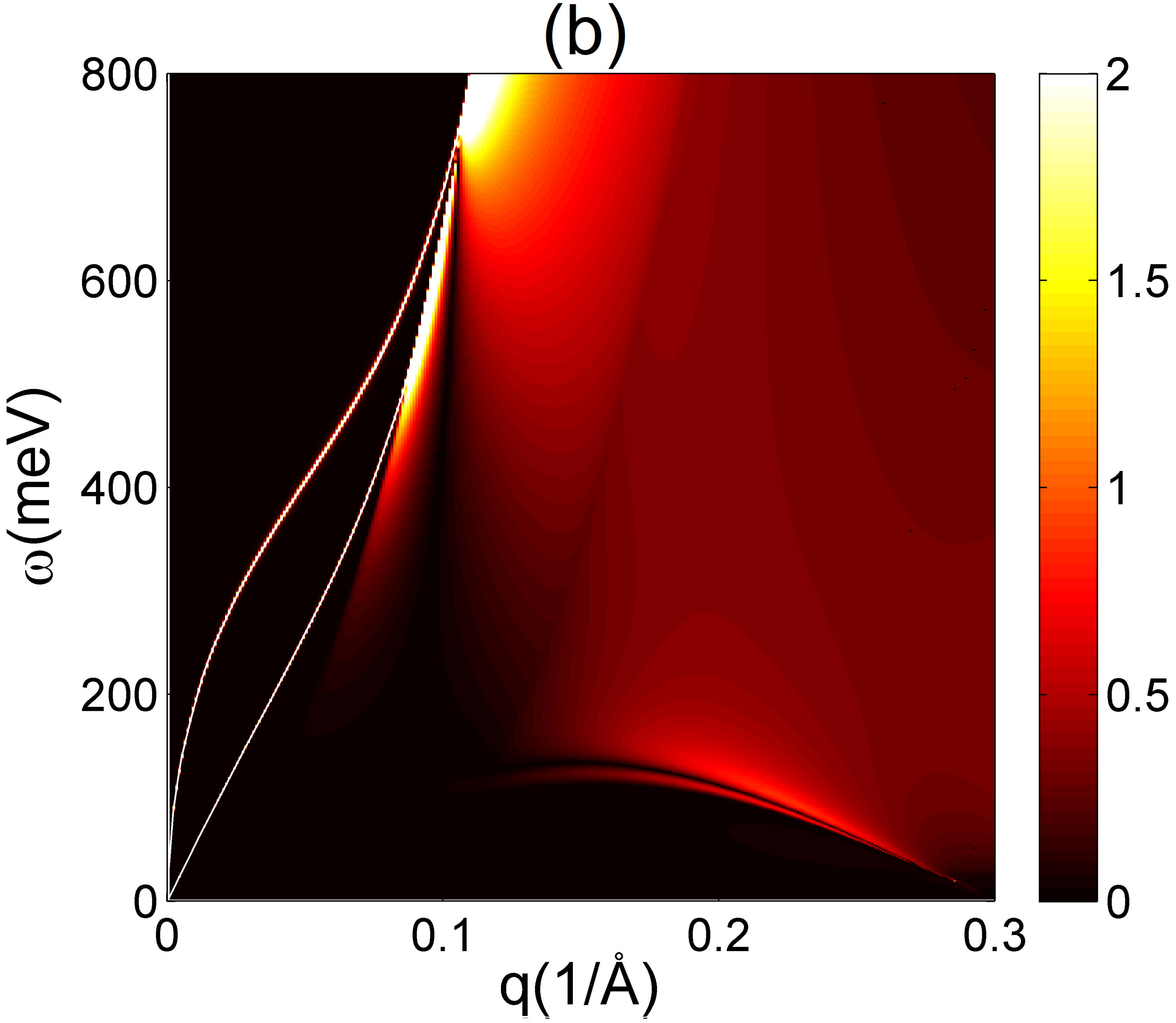}
	\includegraphics[width=5.6cm]{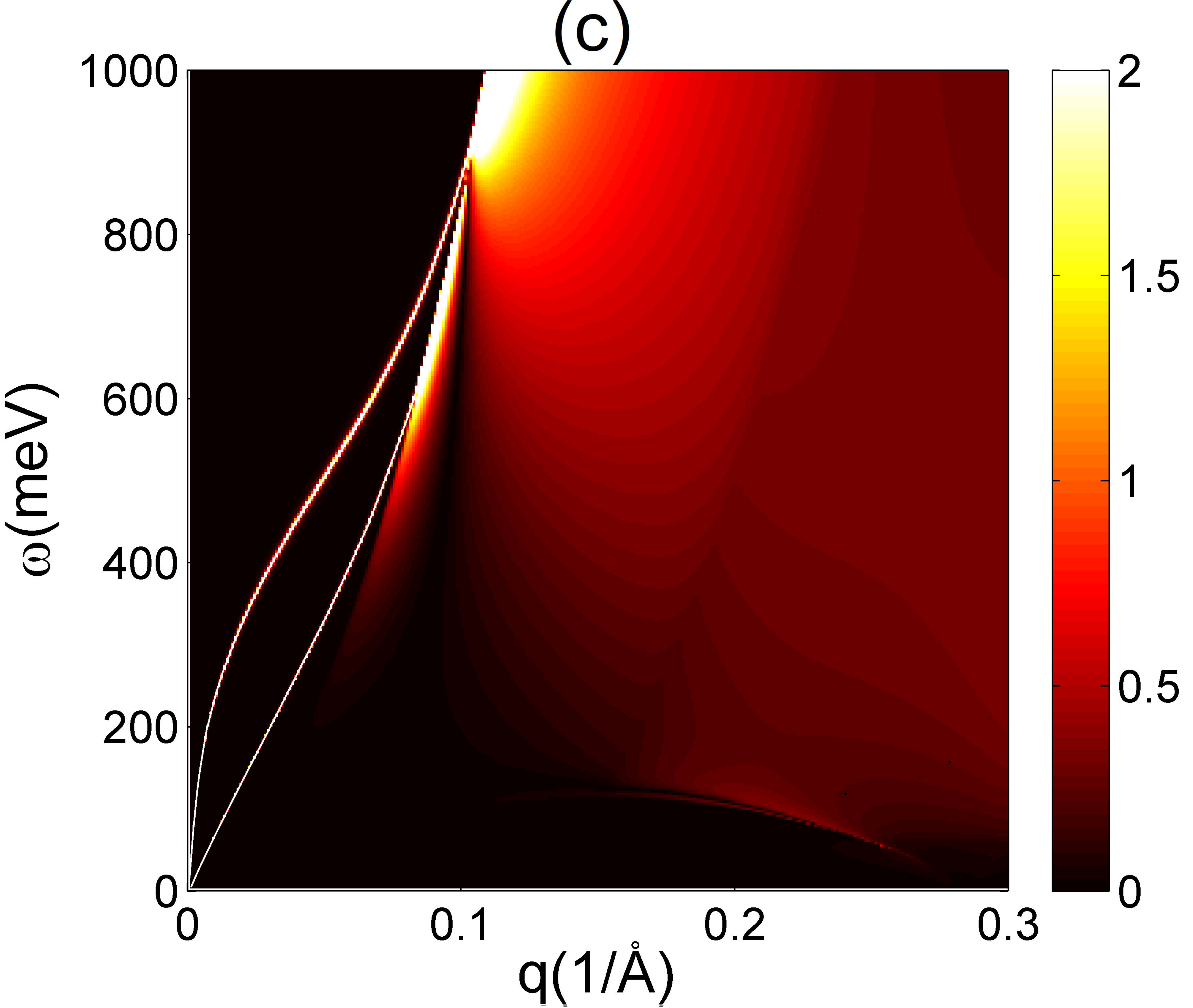}
	\caption{The loss function for the double-layer structure of GaS monolayers at zero temperature and$\ d=15$ {\AA} for various densities: (a)$\ p=2\times10^{13} $cm$^{-2} $ (b)$\ p=4\times10^{13} $cm$^{-2} $ and (c)$\ p=6\times10^{13} $cm$^{-2} $.}\label{Fig-4:loss4}
    \end{figure*}
    \begin{figure*}[h]
	\centering
	\includegraphics[width=5.6cm]{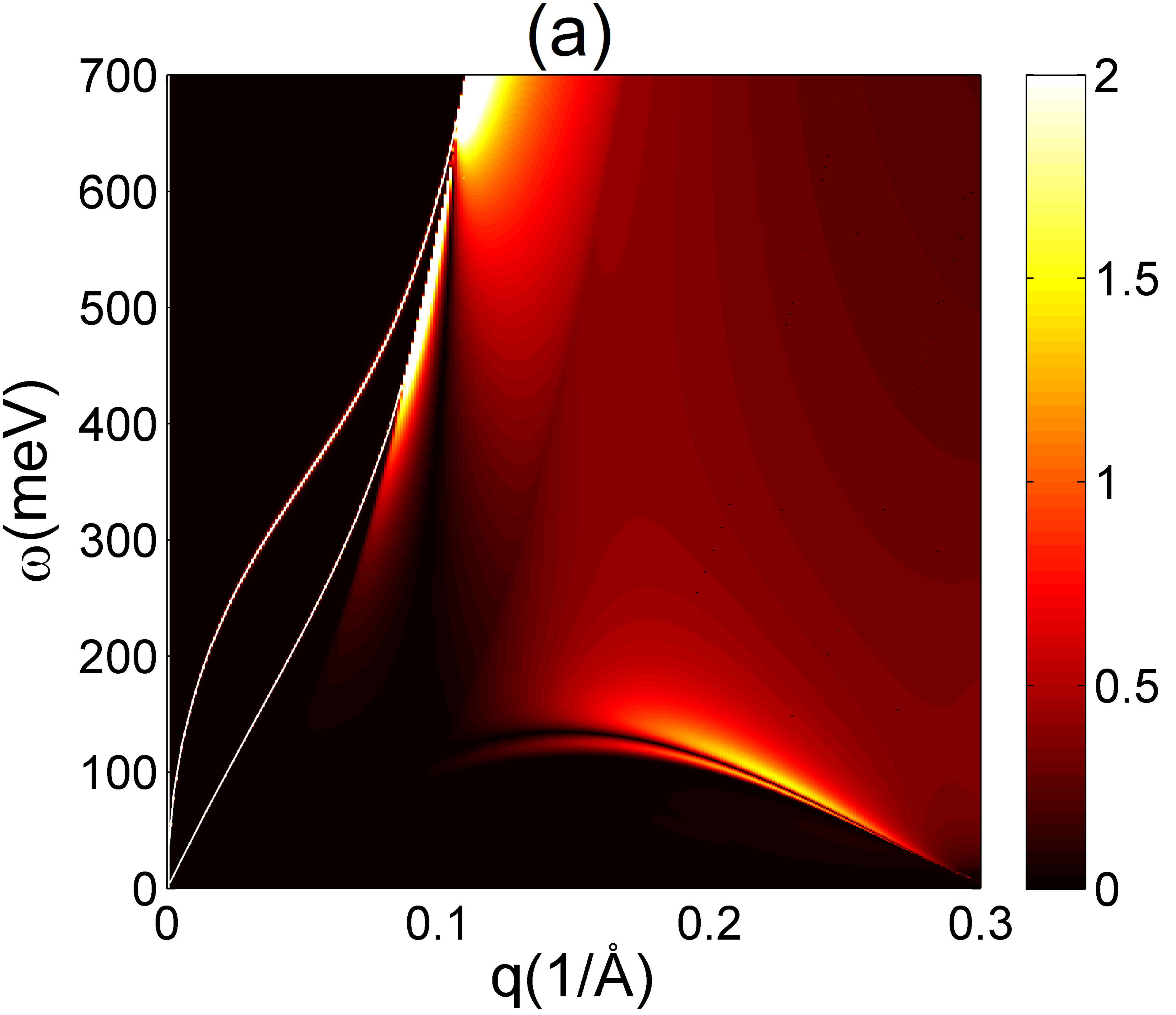}
	\includegraphics[width=5.6cm]{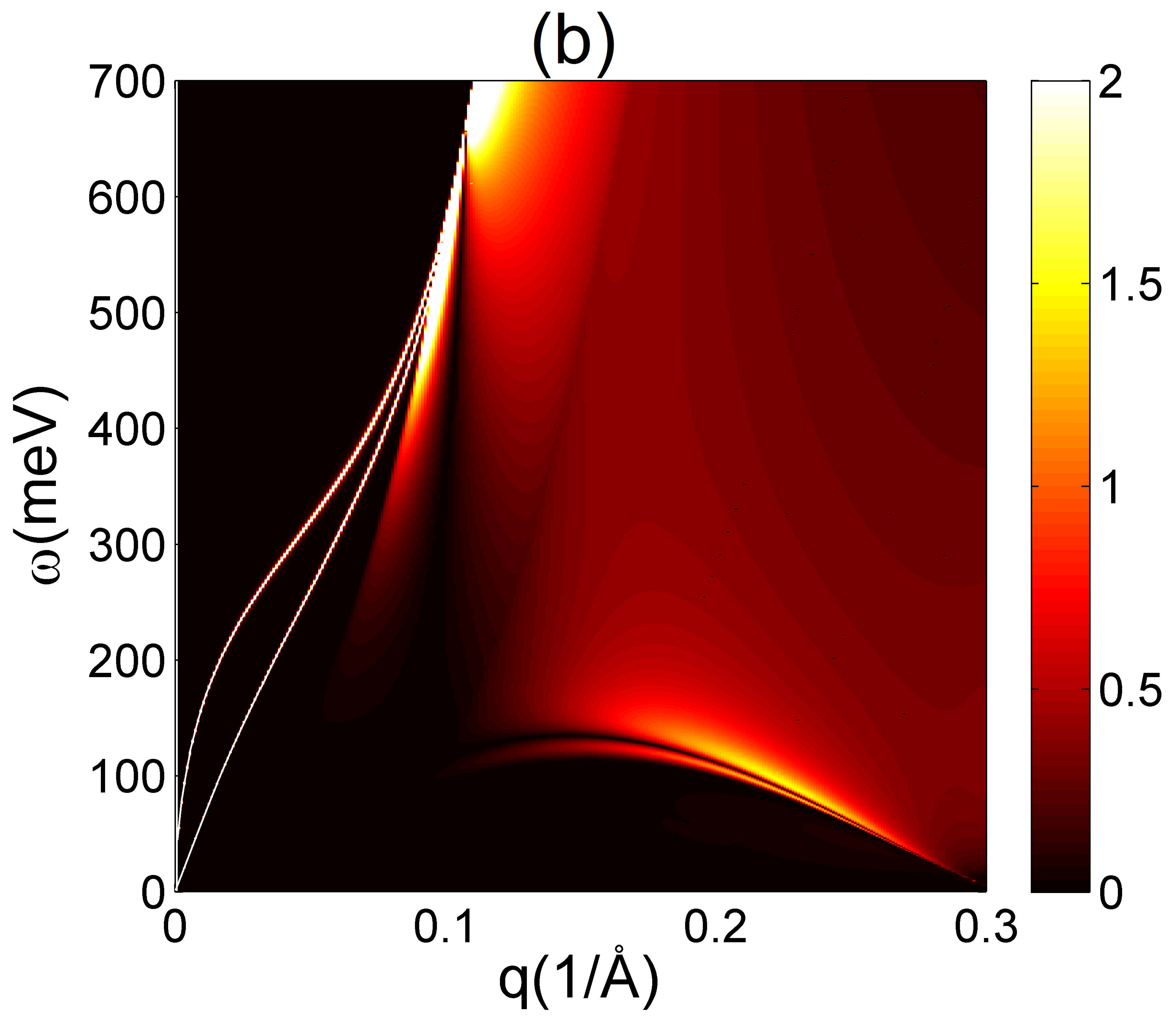}
	\includegraphics[width=5.6cm]{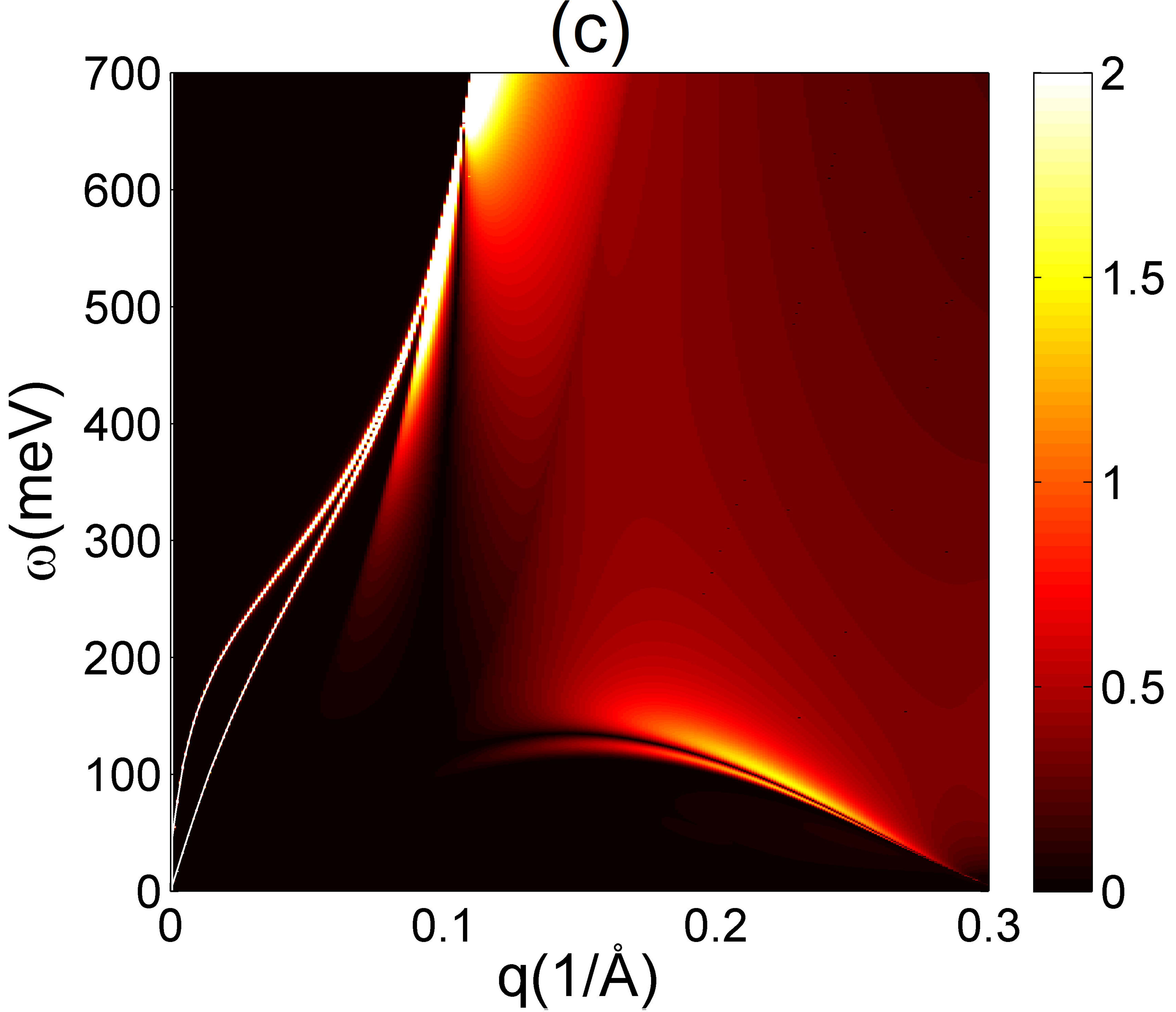}
	\caption{The loss function for the double-layer structure of GaS monolayers at zero temperature and $\ p=3\times10^{13} $cm$^{-2} $ for various layer separations: (a)$\ d=15$ {\AA}, (b)$\ d=30$ {\AA} and (c)$\ d=45$ {\AA}.}\label{Fig-5:loss5}
    \end{figure*}	
	\begin{figure*}[ht!]
	\centering
	\includegraphics[width=7cm]{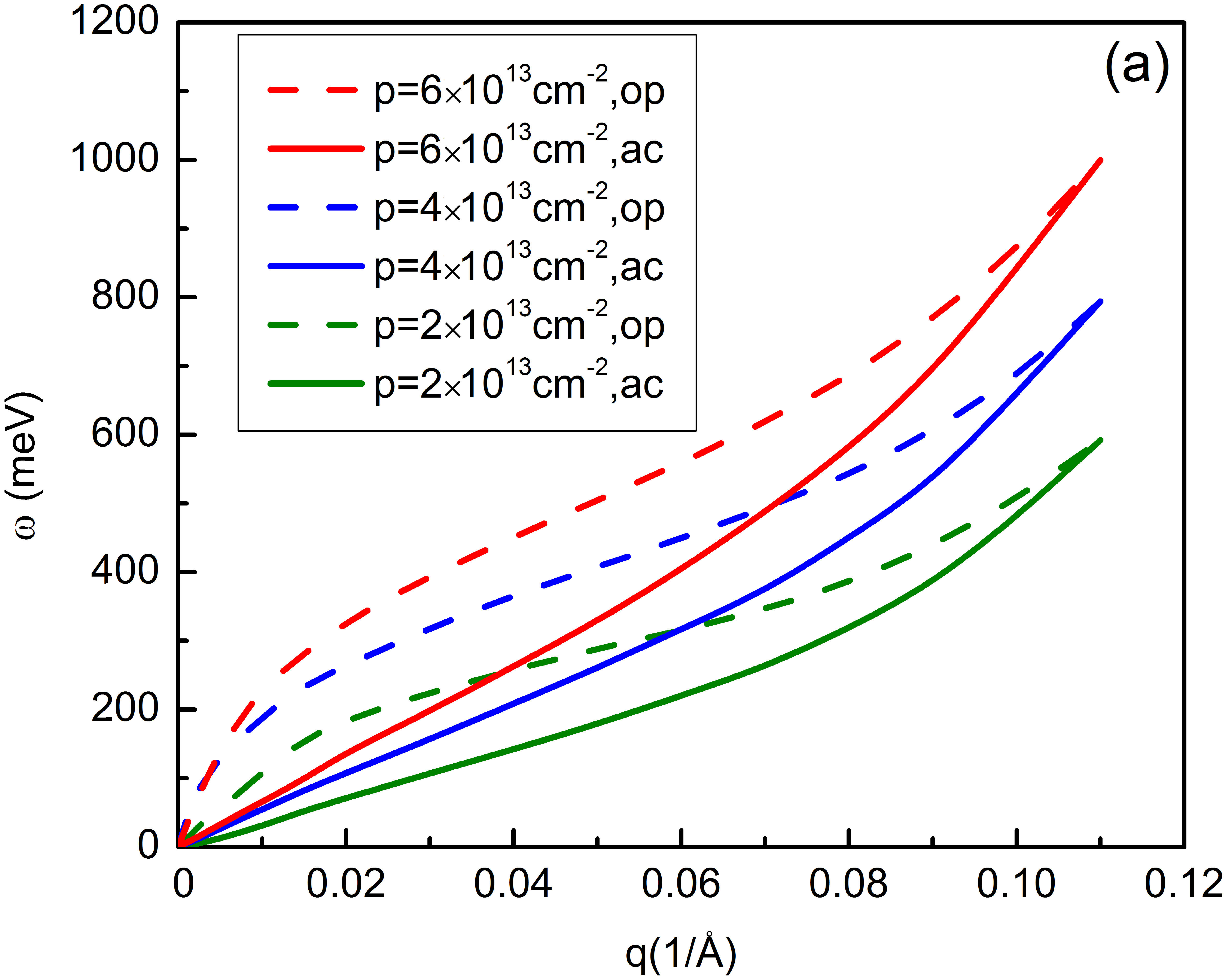}
	\hspace{0.5cm}
	\includegraphics[width=7cm]{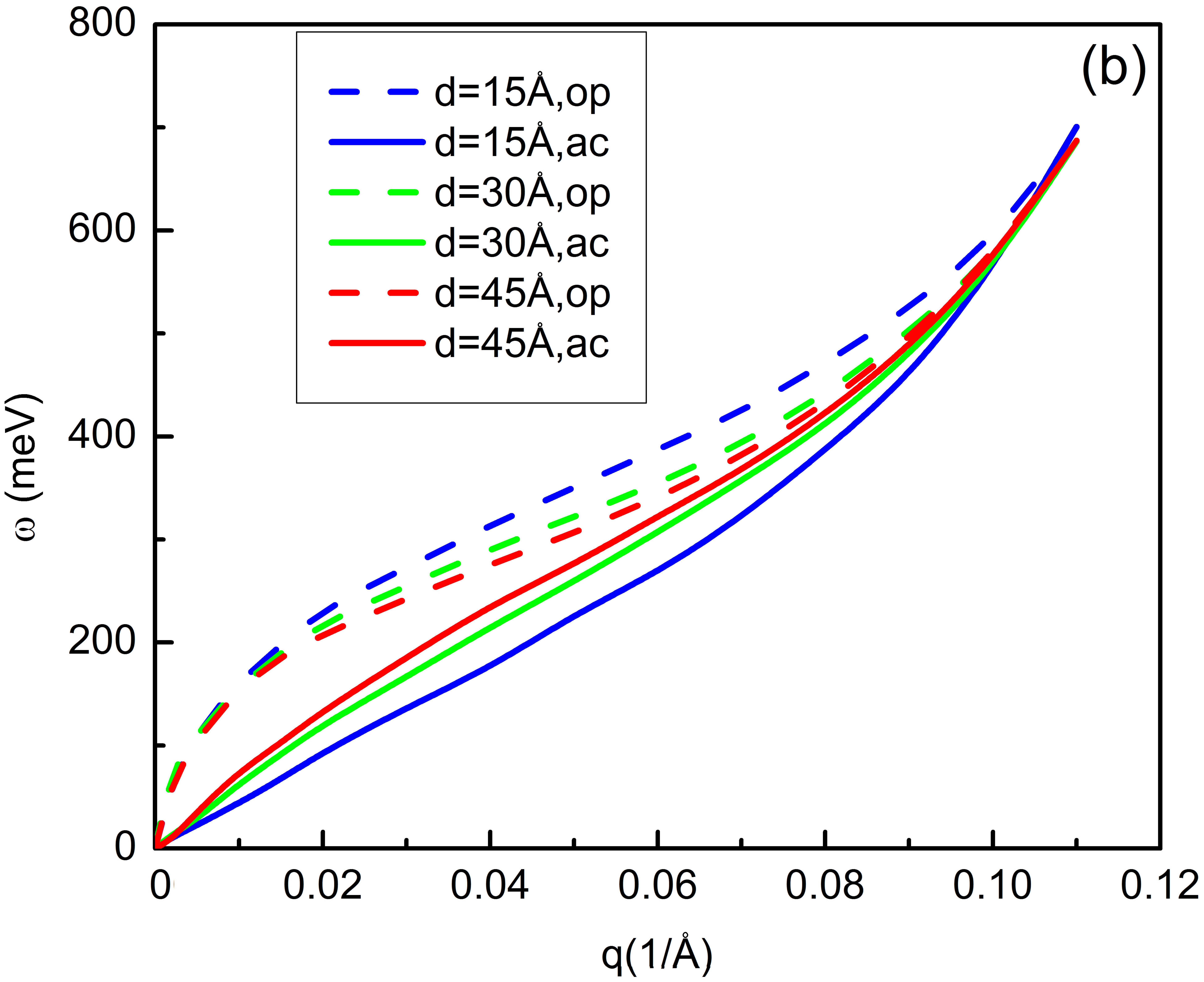}
	\caption{Acoustic and optical plasmon modes behavior at zero temperature for (a) various$\ p=2, 4$ and $\ 6\times10^{13} $cm$^{-2} $and $\ d=15$ {\AA} and (b) various$\ d=15, 30$ and$\ 45${\AA} and $\ p=3\times10^{13} $ cm$^{-2}$.}\label{Fig-6:pl}
	\end{figure*}
	The many-body interaction is taken into account through the dynamically screened Coulomb potential \cite{hwang2007dielectric}:
	\begin{equation}\label{eq:W12}
	W_{12}(\textbf{q},\omega)=\frac{2\pi e^2\ exp({-qd})}{\kappa q \ \varepsilon(q,\omega)}
	\end{equation}
	where$\ d$ is the distance between the two layers,$\ \kappa$ refers to the relative background permittivity and$\ \varepsilon(q,\omega)$ is the dynamical dielectric function. The random phase approximation (RPA) which has been successfully employed for calculating the dielectric function in a double-layer system with identical background permittivity is given by \cite{tanatar2001dynamic}:
	\begin{equation}\label{eq:epsilon}
	\begin{split}
	\varepsilon(q,\omega)=\left(1-\frac{2\pi e^2}{\kappa q}\Pi_{1}(q,\omega)\right)\left(1-\frac{2\pi e^2}{\kappa q}\Pi_{2}(q,\omega)\right)\\
	-\left(\frac{2\pi e^2\ exp({-qd})}{\kappa q}\right)^{2}\Pi_{1}(q,\omega)\Pi_{2}(q,\omega)
	\end{split}
	\end{equation}
	with$\ \Pi_{i}(q,\omega)$ being the 2D non-interacting polarizability of layer$\ i$ at finite temperature: \cite{stern1967polarizability,maldague1978many}
	\begin{equation}\label{eq:Pi}
	\Pi_{i}(q,\omega)=g\sum_{\textbf{k}}\dfrac{f_{i}(\textbf{k})-f_{i}(\textbf{k}{'})}{\Delta E_{\textbf{k},\textbf{k}{'}}+\omega+i\eta^{+}}
	\end{equation}
	In 2D double-layer structures, the collective density fluctuations (plasmons) play an important role in determining the many-body properties of the system such as screening and the drag effect \cite{liu2008plasmon,van2013plasmon}. The plasmon modes are given by the poles of the density-density response function, or equivalently by
	the zeros of the dynamical dielectric function , Eq.(\ref{eq:epsilon}). The loss function, given by$\ -Im[\varepsilon(q,\omega)^{-1}]$, can be used to study the plasmon dispersion,$\  \omega_{p}(q)$.  A plasmon mode appears when both$\ Re[\varepsilon(q,\omega)]$ and$\ Im[\varepsilon(q,\omega)]$ become zero; a situation where$\ -Im[\varepsilon(q,\omega)^{-1}]$ is a $\ \delta$-function with the strength$\ W(q) =\pi[\partial Re[\varepsilon(q,\omega)]/ \partial\omega|_{\omega=\omega_p(q)}]^{-1}$. We start presenting our results with Figure \ref{Fig-3:loss3} where the loss function has been calculated for a p-type doped GaS double-layer structure. We have used$\ m^{*}=0.409 \ m_{0}$,$\ \kappa=3.1$ and$\ E_{0}=111.2 $ meV for GaS \cite{wickramaratne2015electronic,das2019charged}. 
    In Figure \ref{Fig-3:loss3}, we illustrate the loss function in the$\ (q,\omega)$ space at two temperatures,$\ T=0$ and$\ 0.5T_{F}$ for a hole  density $\ p= 4  \times10^{13} $ cm$^{-2}$ and an inter-layer separation of$\ d=15${\AA}. The color scale represents the mode spectral strength. As can be seen from this figure, the single-particle excitations (SPE) continuum has a gap in its low energy part similar to that obtained for a 1D electron gas system (quantum wire). It seems the van Hove singularity in the density of states at band edge which diverges as $\ 1/\sqrt{E}$ is responsible for this newly emerged gap in SPE region of the 2D materials with Mexican hat dispersion. As shown in Figure \ref{Fig-2:mx}(a) there are two Fermi wave vectors$\ {{k}_{F}}_{1}$ and$\ {{k}_{F}}_{2}$ for positive Fermi energies smaller than$\ E_{0}$. They cause the appearance of a narrow SPE band located just below the main dome of SPE continuum (see Figures \ref{Fig-3:loss3}(a) and (b)). The curves in Figure \ref{Fig-3:loss3} indicate the optical and acoustic plasmonic branches and it is notable that the optical branch appears in higher energies. In the acoustic (optical) mode the carriers residing on the two layers oscillate out-of-phase (in-phase), collectively. A comparison between Figures \ref{Fig-3:loss3}(a) and (b) makes it clear that the effect of the finite temperature is to intensify the plasmon damping process. Since at finite temperature hole carriers with larger kinetic energies are excited at negligible energy cost, they enter into the SPE region easier. In Figure \ref{Fig-4:loss4}, we show increasing the carrier density results in shifting the damped optical and acoustic plasmon modes up to higher energies where they eventually enter into the SPE region. Damped plasmons correspond to the broadened peaks in the loss function. Our results show that the density dependence of plasmon modes can be approximated by$\ p^{0.5}$ which happens to be the same behavior as a conventional 2D system with the parabolic energy dispersion \cite{narozhny2016coulomb}.
	In Figure \ref{Fig-5:loss5}, we show the effect of increasing distance$\ (d)$ between the layers on the plasmon modes behavior. We have plotted the loss function for several separations$\ (d=15$,$\ 30$ and$\ 45$ {\AA}), at zero temperature and fixed density$\ (p=3\times10^{13}$ cm$^{-2})$. Calculations indicate that by moving layers away from each other, the optical and acoustic plasmon branches converge and that the mode damping occurs at smaller energies (see Figures \ref{Fig-5:loss5}(a)-(c)). This observation can be attributed to the fact that the inter-layer interaction reduces by increasing inter-layer separation and eventually the system can be considered as two separate layers for which the plasmon branches are degenerate. For a better comparison, the variations of both acoustic and optical plasmon modes with carrier density and layer separation at zero temperature are shown in Figures \ref{Fig-6:pl}(a) and (b), respectively. According to the plasmon branches given in Figure \ref{Fig-6:pl}, we learn that in the limit of long wavelength, the acoustic (optical) plasmon modes show a $\ q  (\sqrt{q})$ dependence in this system which is quite similar to other double-layer structures consisting of 2D materials such as the 2D electron gas, graphene, bilayer graphene, etc. \cite{hwang2011coulomb,hwang2007dielectric,flensberg1995plasmon}. At larger wave vectors, however, the acoustic (optical) plasmon branches have $\ q^{1.3}(q)$ dispersions. The calculations suggest a$\ d^{0.2}$ and a$\ d^{0.1}$ dependency for the acoustic and optical plasmon energies, respectively. 
	\begin{figure}[!ht]
	\includegraphics[width=8cm]{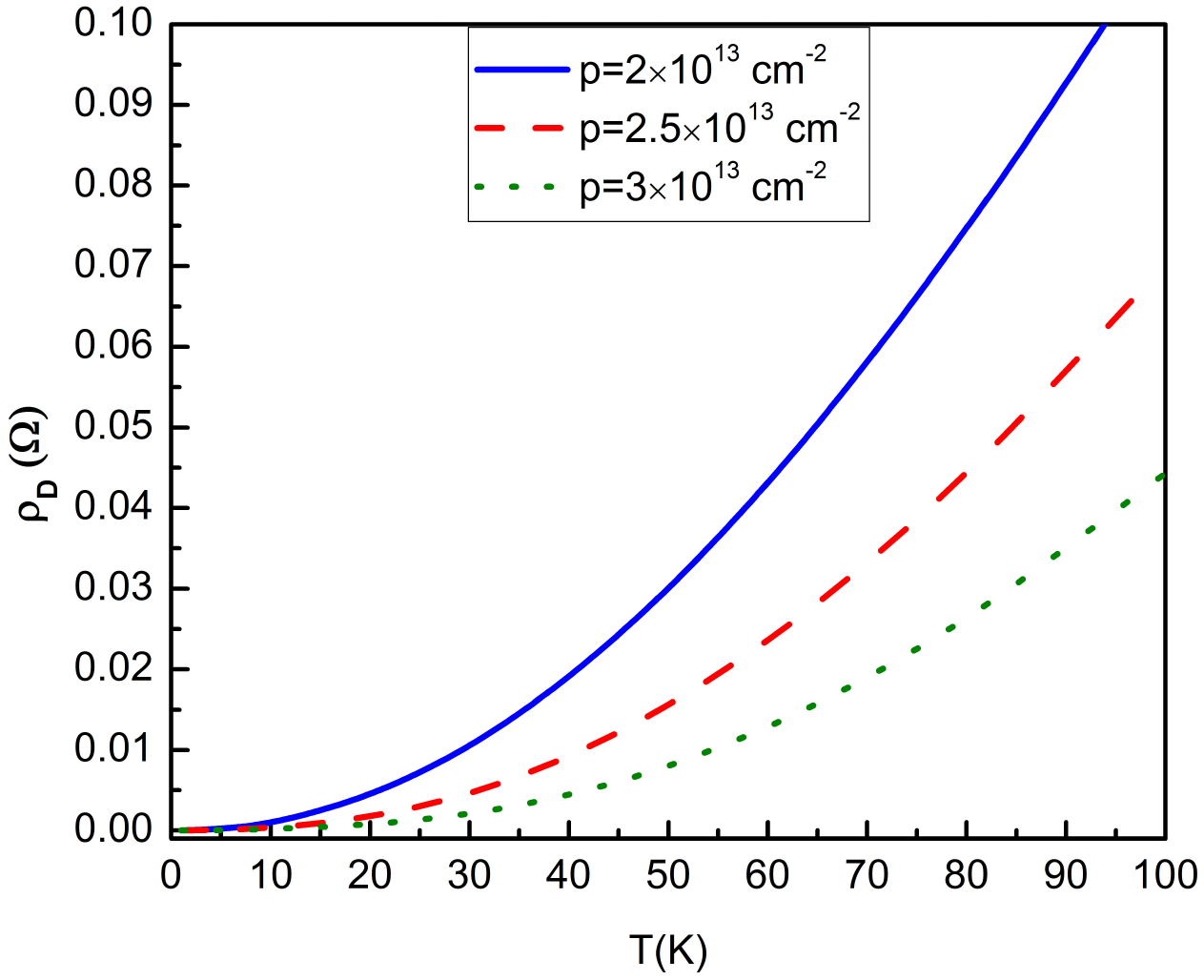} 
	\caption{Drag resistivity as a function of temperature for various densities $ p=2$,$  2.5 $ and $  3 \times10^{13} $ cm$^{-2} $ with $ d=15$ {\AA}.}
	\label{Fig-7:d1}
\end{figure}

\begin{figure}[!ht]
		\includegraphics[width=8cm]{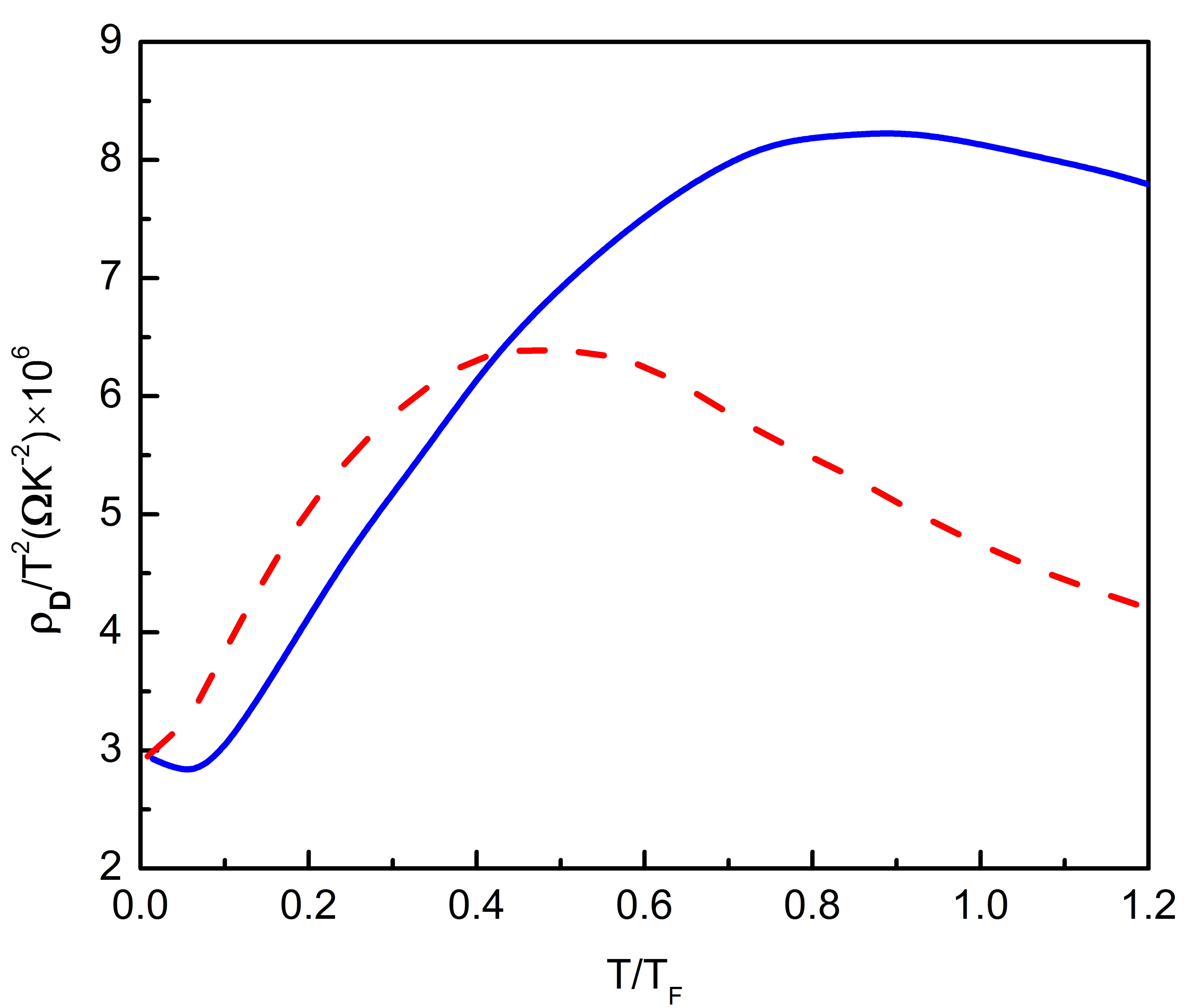} 
		\caption{Drag resistivity scaled by$\ T^{2}$ as a function of temperature at$\ p=2.5 \times10^{13} $ cm$^{-2} $and$\ d=15${\AA}. The solid (dashed) curve shows the corresponding dynamic (static) screening results.}\label{Fig-8:sd}
	\end{figure}

	Now that the effects of$\ T$,$\ p$ and$\ d$ on the plasmon modes of the double-layer system of GaS (as a synthesized 2D material with the Mexican-hat dispersion) are known, we may investigate the Coulomb drag resistivity in such double-layer structure. The calculated drag resistivity as a function of temperature for various densities$\ (p=2, 2.5$ and$\ 3\times10^{13} $ cm$^{-2}) $ at a fixed distance$\ (d=15$ {\AA}) has been shown in Figure \ref{Fig-7:d1}. It can be observed that the drag resistivity decreases with increasing carrier density, at any temperature. To understand this behavior one may note that the plasmon modes take higher energies at higher densities and as a result, they enter into the SPE region easier and get damped faster (see Figures \ref{Fig-4:loss4}(a)-(c)). Therefore, their contribution to the drag resistivity gets weaker and consequently the drag resistivity decreases. One can also learn from Figure \ref{Fig-7:d1} that the drag resistivity rises when the temperature increases at a constant density. Eq. (\ref{eq:rhoD}) can explain this observation: there are two types of important contributions to the Coulomb drag resistivity;$\ Im(\Gamma_{i}(\textbf{q},\omega))$ and$\ W_{12}(\textbf{q},\omega)$. At zero temperature, the well-defined plasmon modes always lie outside the SPE region and there is no coupling between SPE region and plasmon modes$\ (Im(\Gamma_{i}(\textbf{q},\omega)=0)$ which results in$\ \rho_{D}=0$. It is obvious in Figures \ref{Fig-3:loss3}(a) and (b), that by increasing the temperature, the SPE continuum and plasmon peaks are broadened and partially overlapped due to the thermally activated holes. In this situation$\ Im(\Gamma_{i}(\textbf{q},\omega))$ has a non-zero value, resulting in the plasmon contributions enhancement (described by the zeros of the dielectric function$\ \varepsilon(q,\omega)$) to$\ \rho_{D}$.
	On the other hand, according to our calculations which have been performed for several hole densities and inter-layer separations, we have found that the temperature dependence of the drag resistivity can be approximated as$\ T^{2}$ at low temperature and for$\ k_{F}d>1$. This behavior has been reported for other double-layer Fermi systems like double-quantum well with parabolic energy dispersion \cite{jauho1993coulomb}. At intermediate temperatures, a$\ T^{2.8}$ dependence has been obtained for GaS which is due to the plasmon enhancement effect. This effect can be clearly observed in  Figure \ref{Fig-8:sd} where both the statically and dynamically screened results of the drag resistivity (scaled by$\ T^{2}$) have been shown as a function of temperature. The calculations for a hole density of $\ 2.5 \times10^{13} $ cm$^{-2} $ suggest that the plasmon contribution to the drag resistivity becomes important as $T$ increases above an intermediate temperature $\sim 0.45T_{F}$ and exhibits a peak around $T= 0.9T_{F}$. 	
	\begin{figure}[!h]
		\includegraphics[width=8cm]{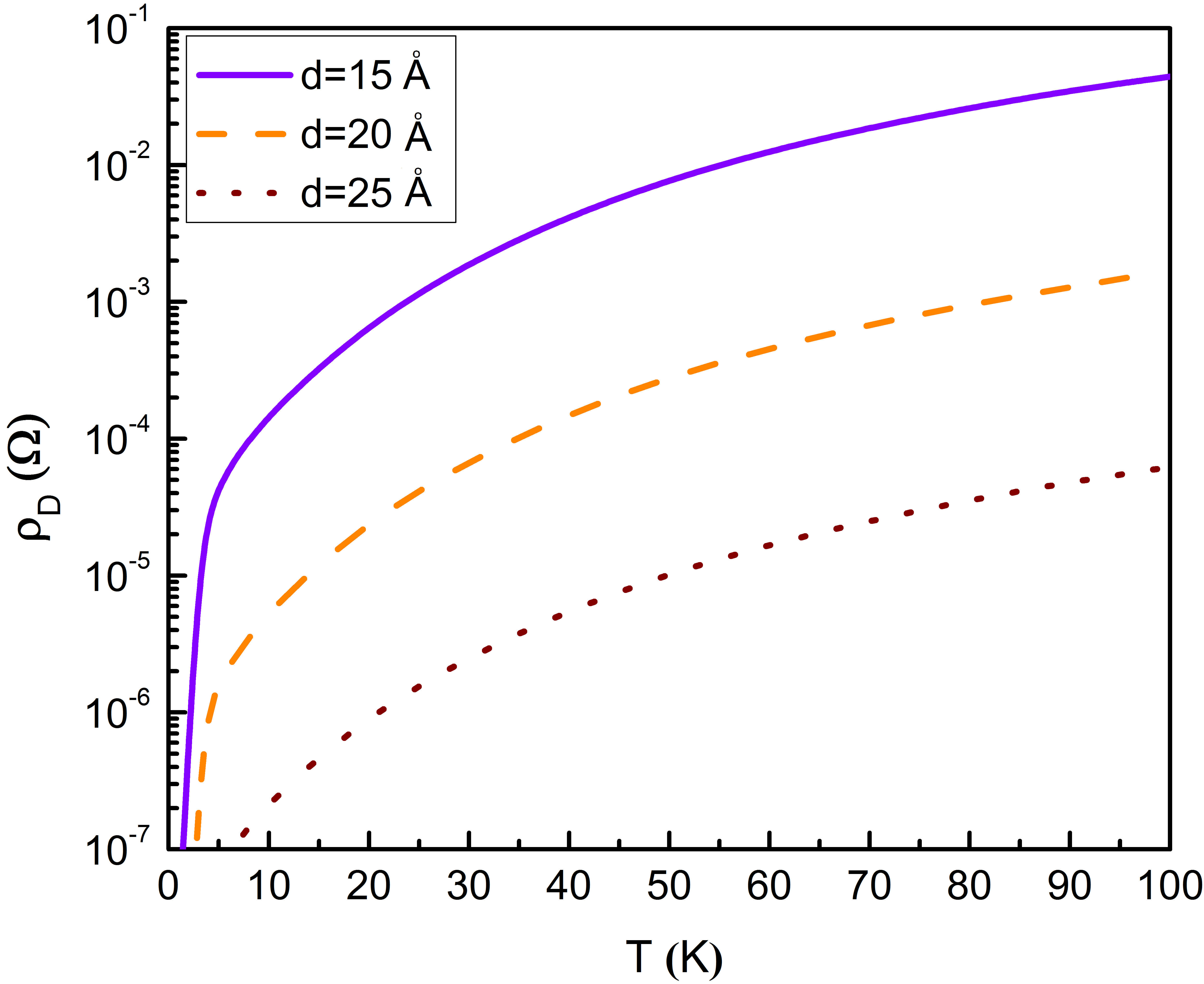}
		\caption{Coulomb drag resistivity as a function of temperature for various layer separations$\ d=15$ {\AA},$\ 20$ {\AA} and$\ 25$ {\AA} at$\ p=3\times10^{13} $ cm$^{-2}$.}
		\label{Fig-9:d9}
	\end{figure}
	\begin{figure}[!h]
		\includegraphics[width=8cm]{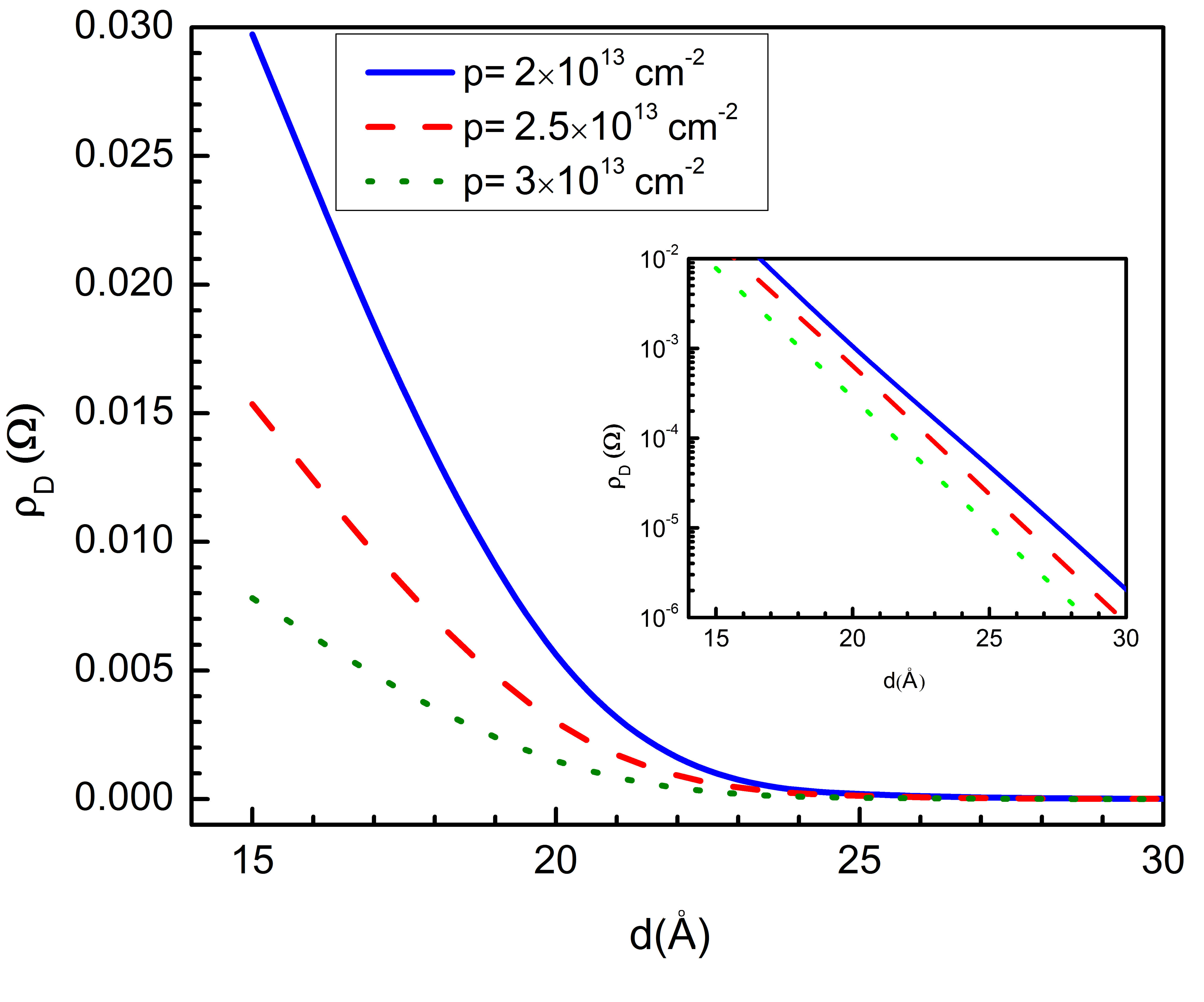}
		\caption{Inter-layer separation dependence of the Coulomb drag resistivity in a double-layer GaS at$\ T=50$ K and for $\ p=2, 2.5 $ and $3\times10^{13} $ cm$^{-2}$.}
		\label{Fig-10:d10}
	\end{figure}
	\begin{figure}[!h]
		\includegraphics[width=8cm]{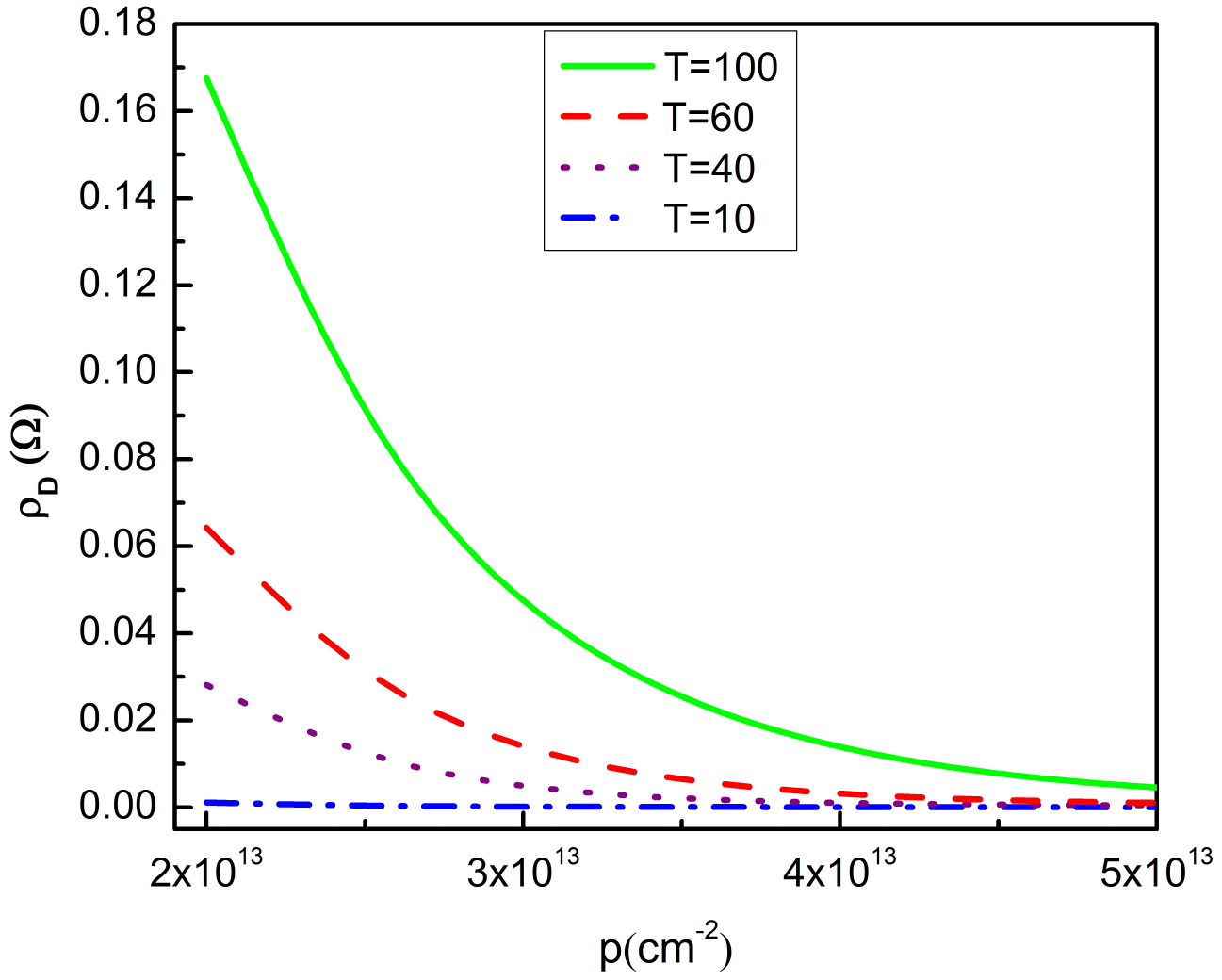}
		\caption{Density dependence of the Coulomb drag resistivity in a double-layer GaS system at$\ T=10,40,60 $ and$ 100$ K with$\ d=15$ {\AA}.}
		\label{Fig-11:d11}
	\end{figure}
	\begin{figure*}[ht!]
		\centering
		\includegraphics[width=7cm]{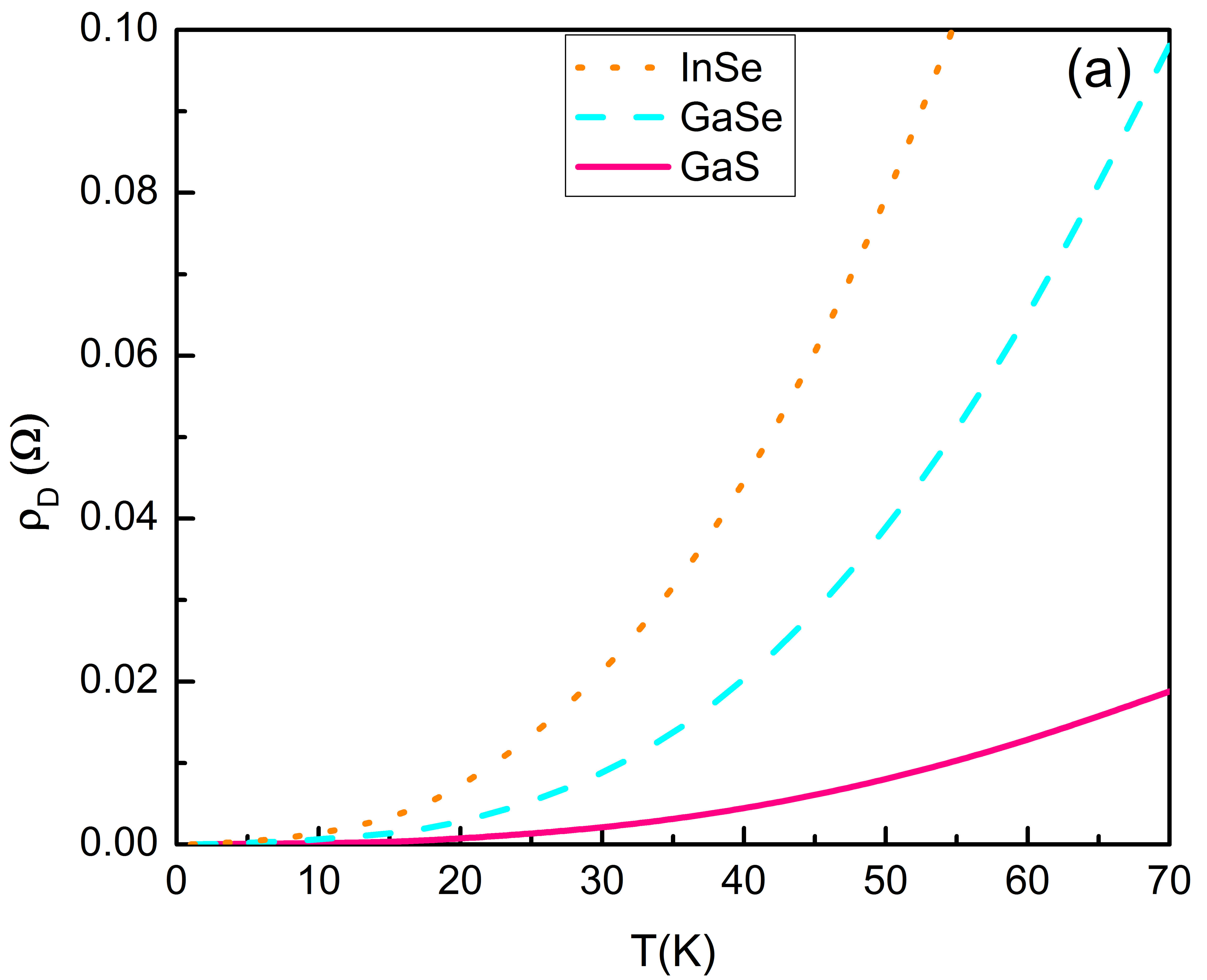}
		\hspace{0.5cm}
		\includegraphics[width=7cm]{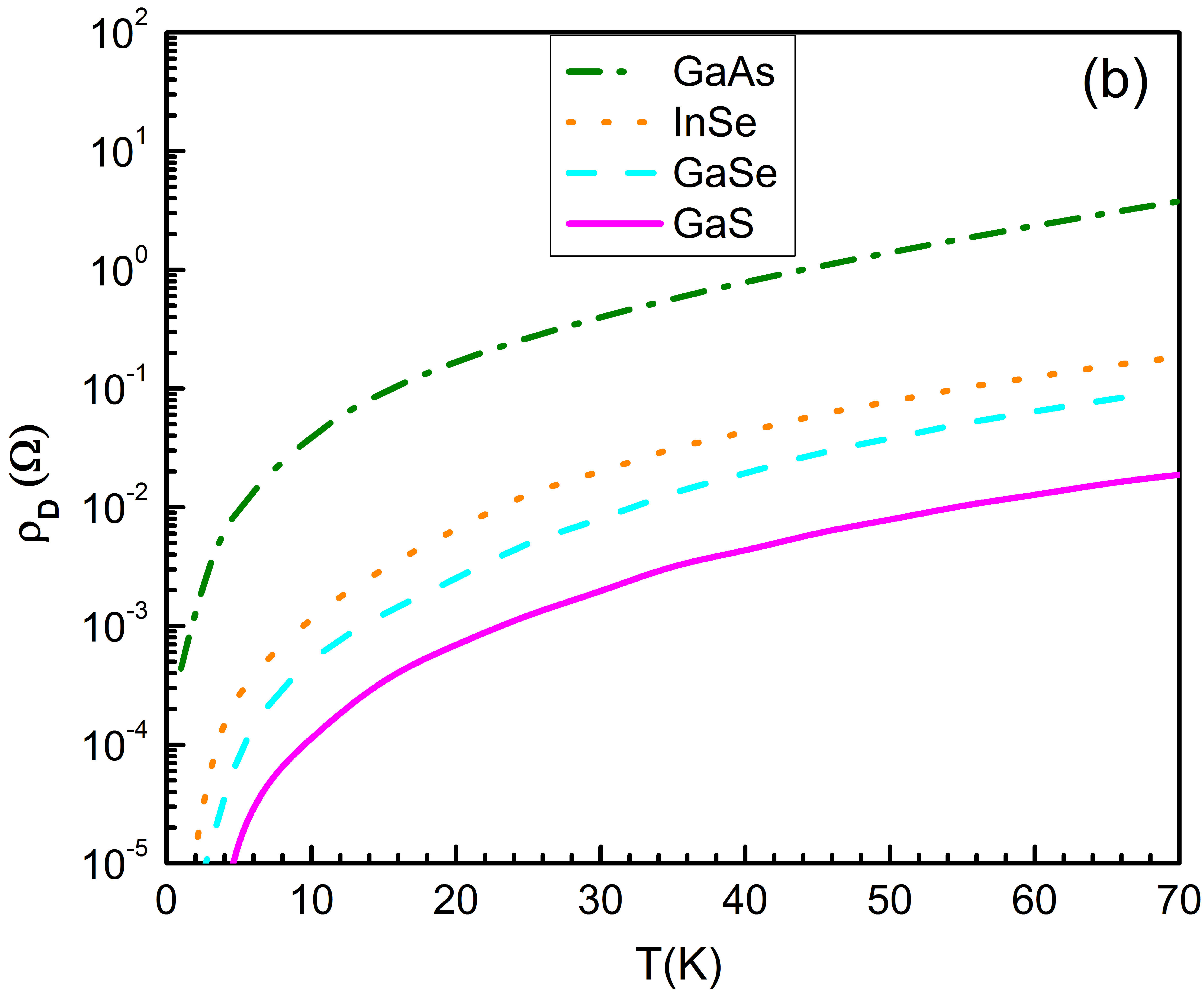}
		\caption{Coulomb drag resistivity as a function of temperature in (a) the double-layer of GaS, GaSe and InSe systems at $\ p=3\times10^{13} $ cm$^{-2}$ and$\ d=15$ {\AA} with $\ k_{F}d\sim 2-3$   and (b) a double-layer of GaAs-based 2D electron gas with $\ k_{F}d= 2.44$, compared to those given in (a).}\label{Fig-12:d12}
	\end{figure*}
	As we mentioned before, the effect of the inter-layer spacing,$\ d$, on the drag transresistivity is also of interest. In Figure \ref{Fig-9:d9}, we have presented calculations for drag resistivity as a function of temperature for three layer separations$\ (d=15$, $\ 20$ and$\ 25$ {\AA}) at a fixed density$\ (p=3\times10^{13}$ cm$^{-2})$. Our results suggest that$\ \rho_{D}$ decreases with increasing$\ d$. It is not surprising though, because by increasing$\ d$ the Coulomb interaction between layers decreases and consequently the inter-layer coupling becomes weaker. This is obvious that no Coulomb drag effect appears when the two layers are sufficiently far away.
	In addition, Figure \ref{Fig-10:d10} demonstrates the drag resistivity as a function of distance between centers of layers at $\ T=50$ K and for three different hole densities, $\ p=2, 2.5 $ and $3\times10^{13} $ cm$^{-2} $. As it can be observed, $\ \rho_{D}$ reduces exponentially with increasing the layers separation for all hole densities. Interestingly, this behavior has been obtained for a double-quantum wire system, experimentally \cite{debray2001experimental}.	
	To illustrate the behavior of the drag transresistivity more clearly, we have displayed the change of$\ \rho_{D}$ with the hole density at four different temperatures,$\ T=10,40,60 $ and $ 100 $ K in Figure \ref{Fig-11:d11}.  Calculations show that the density dependence of the drag resistivity varies with temperature and it can approximately be given as$\ p^{-4}$ ($\ p^{-4.5}$) at low (intermediate) temperatures.
		
	Now we are all set to step forward and look into other important materials in the same family as GaS. In Figure \ref{Fig-12:d12}(a), we have compared the temperature dependence of the drag resistivity in the case of double-layer GaS with those obtained for double-layer GaSe and double-layer InSe systems at a fixed density and layer separation. It should be pointed out that a different set of parameters, including effective mass, relative permittivity and the Mexican-hat height, defines each of the mentioned materials. Here, the corresponding parameters for GaSe are$\ m^{*}=0.6 m_{0}$,$\ \kappa=3.55$ and$\ E_{0}=58.7$ meV and for InS the parameters are $\ m^{*}=0.926 m_{0}$,$ \ \kappa=3.38$ and$\ E_{0}=34.9 $ meV \cite{wickramaratne2015electronic,das2019charged}. The obtained results suggest that the drag transresistivity  decreases with increasing the Mexican-hat height so that GaS with the largest Mexican-hat takes smaller values of drag resistivity at any temperature and it is InSe that provides the highest drag resistivity among the materials studied here.
	In addition,  while$\ \rho_{D}$ shows a$\ T^{2}$ dependency at low temperatures, their drag resistivities at intermediate temperatures have a faster growth with$\ T$ (i.e. $\ T^{2.8}$) for all
	double-layer systems studied here. It occurs because of  enhancing contributions of the plasmon modes in drag resistivity.  In Figure \ref{Fig-12:d12}(b), we compare our results shown in Figure \ref{Fig-12:d12}(a) with a system consisting of two parallel layers of GaAs-based 2D electron gas (quantum well) with parabolic energy dispersion. We set the value of $\ k_{F}d=2.44$ for this 2D electron gas system close to the values we used in Figure \ref{Fig-12:d12}(a), $\ k_{F}d \sim 2-3$, to make sure all systems to be in the same coupling regime. It should be noted that the carrier densities and Fermi energies do not match in this comparison. The results suggest that the drag resistivity of the parabolic dispersion system takes higher values than those in our Mexican-hat dispersion systems. It seems the differences in the SPE regions of the two systems could probably account for this observation; the opening of a gap in the SPE continuum reduces the contribution of $\ Im\Pi\neq0$ to the drag resistivity in our system compared to the conventional 2D electron gas. \\
	\section{Conclusion}
	To summarize, first we have investigated the behavior of plasmons and SPE region in the double-layer system with Mexican-hat bandstructure which consists of two p-type doped GaS monolayers in close proximity with no tunnelings. Our numerical results show that the damped optical and acoustic plasmon branches shift to higher energies and then enter into the SPE region when density is increased. In addition, the density dependence of plasmon modes is approximately$\ p^{0.5}$. Moreover, at fixed density and finite temperature, plasmon modes damping accelerates in comparison with that at zero temperature. Besides, we have found that the dependence of acoustic and optical modes to the inter-layer spacing can be approximated as$\ d^{0.2}$ and$\ d^{0.1}$, respectively. 
		Also, the acoustic (optical) plasmon branch follows a $\ q$ ($\ \sqrt{q}$) dispersion at long wavelengths and shows a $\ q^{1.3}$ ($\ q$) behavior at larger wave vectors, before entering the SPE damping region. According to our results, the drag resistivity has a temperature dependence as$\ T^{2}$ ($\ T^{2.8}$) at low (intermediate) temperatures. Our calculations also show the drag resistivity decreases exponentially with increasing layers separation similar to the case of double-quantum wire system. Furthermore, we note that although the change of transresistivity with the hole density can be approximated as$\ p^{-4}$ at low temperatures, it exhibits a faster reduction at higher temperatures. It has been found that the change of $\ \rho_{D}$ with the carrier density (layer separation) follows the same behavior as in double-quantum well (double-quantum wire) system. 
	Finally, we have compared the temperature dependence of the drag resistivity for three materials with Mexican-hat valence band dispersion (GaS, GaSe and InSe) and shown the drag resistivity value of GaS is the smallest while its Mexican-hat is the largest.
	
\end{document}